\documentclass[fleqn,usenatbib]{mnras}

\usepackage{newtxtext,newtxmath}
\usepackage[T1]{fontenc}

\usepackage{booktabs}

\DeclareRobustCommand{\VAN}[3]{#2}
\let\VANthebibliography\thebibliography
\def\thebibliography{\DeclareRobustCommand{\VAN}[3]{##3}\VANthebibliography}

\usepackage{graphicx}	% Including figure files
\usepackage{amsmath}	% Advanced maths commands

\title[Formation of ultra-diffuse galaxies via mergers]{Formation of globular cluster-rich ultra-diffuse galaxies through mergers}

\author[T. Tapia et al.]{
Truman Tapia,$^{1}$\thanks{E-mail: truman.tapiamora@icrar.org}
Hidenori Matsui,$^{1,2}$
Kenji Bekki,$^{1}$
O. Ivy Wong,$^{1,3}$
Connor Bottrell,$^{1}$
Brent Groves,$^{1}$
\newauthor
 Duncan A. Forbes,$^{4}$
Jean P. Brodie,$^{4,5}$
Warrick J. Couch$^{4}$
\\
$^{1}$International Centre for Radio Astronomy Research, The University of Western Australia, 35 Stirling Highway, Crawley, Western Australia 6009, Australia\\
$^{2}$National Institute of Technology, Asahikawa College, Shunkodai 2-2-1-6, Asahikawa, Hokkaido, 071-8142, Japan\\
$^3$ ATNF, CSIRO Space and Astronomy, Bentley, WA, Australia\\
$^4$Centre for Astrophysics and Supercomputing, Swinburne University, John Street, Hawthorn, VIC 3122, Australia\\
$^5$University of California Observatories, 1156 High Street, Santa Cruz, CA 95064, USA
}

\date{Accepted XXX. Received YYY; in original form ZZZ}

\pubyear{\the\year{}}

\begin{document}
\label{firstpage}
\pagerange{\pageref{firstpage}--\pageref{lastpage}}
\maketitle

\begin{abstract}

\noindent We use high-resolution, idealized hydrodynamic simulations of gas-rich dwarf-galaxy mergers to test whether such encounters can form ultra-diffuse galaxies (UDGs) with globular cluster (GC) systems. We simulate 1:1 and 1:2 mergers alongside an isolated control model and identify stellar overdensities as GC candidates (GCCs). The remnants evolve into dispersion-supported, UDG-like systems with three-dimensional stellar half-mass radii $r^{\rm 3D}\sim1.9$--$2.6$ kpc, while the isolated dwarf remains rotationally supported and forms no GCCs. Tidal heating and stellar feedback expel a large fraction of the gas beyond the dark matter (DM) halo, leaving stellar-dominated remnants whose DM haloes remain cuspy. Merger-driven star formation is highly clustered: the fraction of newly formed stellar mass bound in massive clusters exceeds $0.5$ after the first pericentric passage and remains elevated thereafter. By the final snapshot, the remnants host GC populations numbering $20$ (1:1) and $39$ (1:2), more centrally concentrated than the field stars and consistent with the observed GC number--halo mass relation. The GCCs match observed star clusters in the planes of mass versus size, velocity dispersion, and density. More massive clusters exhibit stronger internal rotation and broader metallicity spreads. In one case, the merger produces a nucleated UDG via cluster inspiral followed by sustained in-situ star formation. These results demonstrate that gas-rich dwarf mergers are a viable pathway to GC-rich---and sometimes nucleated---UDGs, and predict correlated cluster mass, rotation, and metallicity-dispersion trends testable with observations.

\end{abstract}


%%%%%%%%%%%%%%%%%%%%%%%%%%%%%%%%%%%%%%%%%%%%%%%%%%

%%%%%%%%%%%%%%%%% BODY OF PAPER %%%%%%%%%%%%%%%%%%

\section{Introduction}

Ultra-diffuse galaxies (UDGs) have garnered significant attention since \citet{van2015forty} reported the discovery of 47 such objects in the Coma cluster\footnote{Galaxies meeting the UDG criteria were recognized earlier (e.g., \citealt{sandage1984studies, reaves1956dwarf, disney1976visibility}).}. Typically defined by a central surface brightness of $\mu(g, 0)\geq 24$ mag arcsec$^{-2}$ and an effective radius $r_e\geq 1.5$ kpc (see \citealt{van2022s} for a discussion on the definition), these galaxies exhibit low surface brightness, extended sizes, and stellar masses comparable to those of dwarf galaxies $M_\star\sim 10^7-10^9$ $\rm M_\odot$ \citep{10.1093/mnras/stx2648, Leisman_2017}. Despite their unusual properties, the formation mechanisms of UDGs remain actively debated.

Observational efforts have identified UDGs in wide-area imaging surveys (e.g. \citealt{2019ApJS..240....1Z, Zaritsky_2022, Zaritsky_2023, 2024ApJ...976...75S}), and across a wide range of environments—from the field \citep{Leisman_2017, papastergis2017hi, bellazzini2017redshift} and galaxy groups \citep{vzemaitis2023tale, makarov20156, roman2017ultra, forbes2019ultra} to rich clusters \citep{van2015forty, mihos2015galaxies, venhola2017fornax, gannon2022ultra}. In clusters, UDGs are typically gas-depleted, red, and quenched, whereas in lower-density environments they tend to be gas-rich, blue, and star-forming \citep{prole2019observational, roman2017ultra, marleau2024dwarf, spekkens2018atomic, scott2021resolved}. Notably, exceptions to these trends exist \citep{papastergis2017hi, kadowaki2017spectroscopy, for2023wallaby}.

A recent study by \citet{buzzo2025multiple} analyzing 88 UDGs (plus 36 near UDGs) identified two distinct classes. One class, predominantly found in the field and associated with high gas content, is characterized by blue colors, young ages, elongated shapes, low surface brightness, extended star formation histories, and GC-poor systems that follow the classical dwarf mass--metallicity relation (MZR). In contrast, a second class—typically residing in denser environments—displays red colors, older ages, rounder shapes, larger sizes, higher surface brightness, shorter star formation timescales, and a large population of GCs, with these galaxies lying below the dwarf MZR. \citet{buzzo2025multiple} interpret these findings as evidence for two formation channels: a “puffy dwarf” origin for the GC-poor systems and a “failed galaxy” origin for the GC-rich systems.

The failed galaxy model, originally proposed by \citet{van2015forty}, posits that UDGs are the quenched remnants of more massive galaxies (e.g., M33) whose stellar populations evolved passively after early quenching—potentially driven by environmental processes such as ram pressure stripping \citep{yozin2015quenching}, tidal interactions \citep{carleton2019formation}, or GC quenching \citep{Danieli_2022}. In this scenario, the high-mass halo of the progenitor already contains a large population of GCs as they form in the early stages of the galaxy and the mass of the GC system is correlated with the halo mass \citep{spitler2009new, burkert2020high, harris2015dark, forbes2018extending, harris2017galactic, forbes2024ultra, forbes2025some}. Alternatively, the puffy-dwarf model suggests that UDGs originate directly from dwarf galaxies through mechanisms including stellar feedback-driven expansion \citep{di2017nihao}, high angular momentum \citep{amorisco2016ultradiffuse}, tidal effects \citep{roman2021diffuse, duc1998young, carleton2019formation}, or mergers \citep{wright2021formation}. Given the complex observational landscape, a hybrid scenario combining multiple formation pathways is likely necessary to explain the wide population \citep{rong2017universe, martin2019formation}.

Intriguingly, individual UDGs with atypical properties challenge these broad classifications. \citet{fielder2023disturbed} reported UGC 9050-Dw1—a gas-rich UDG with $\log M_*/\rm M_\odot\sim7.5$ and $\log \rm M_{HI}/\rm M_\odot\sim8.4$—located on the outskirts of a group and accompanied by a low surface brightness companion. Featuring a central UV-emitting clump and an unusually rich population of 52 monochromatic GCs (contributing approximately 20\% of its total light), this object defies the typical association of GC richness with red, quenched systems in clusters. The authors suggest that a merger is the most likely origin of this galaxy. A similar case is presented by \citet{fielder2024all} with KUG 0203-Dw1, which, despite its gas content and UV emission, hosts only about eight GCs. In this instance, the authors suggest that tidal forces likely transformed a regular dwarf galaxy into a UDG, given that the GC count falls within the expected range for dwarfs. These examples underscore the need for detailed theoretical predictions on the role of mergers in shaping both the structural properties of UDGs and their GC populations.

Although several conceptual models have been proposed—including \citet{silk2019ultra} and \citet{baushev2018galaxy}—the formation of UDGs via mergers remains underexplored. Cosmological simulations offer a promising avenue: for example, \citet{wright2021formation} showed that dwarf–dwarf mergers in ROMULUS25 can produce galaxies with UDG properties by boosting the spin of the remnant. Such encounters can also trigger the formation of massive star clusters (SCs), a process that can be investigated with high-resolution simulations \citep{2025A&A...704A.240D, lahen2019formation, saitoh2009toward, matsui2025formation}. However, this line of inquiry has not yet been pursued explicitly in the context of UDGs.

A growing literature treats GC formation in galaxies and UDG formation, yet the specific problem of GC formation within UDGs remains open. Theoretical studies of GC formation span semi-analytic models \citep{2002MNRAS.333..383B, 2021MNRAS.505.5815V}, constrained galaxy-scale simulations \citep{2002MNRAS.335.1176B}, cosmological simulations \citep{10.1093/mnras/stx3124}, and galaxy-scale simulations that resolve the birth of massive clusters \citep{lahen2019formation, matsui2025formation}. For UDGs with rich GC systems, progress has come from semi-analytic treatments \citep{2021MNRAS.502..398C} and from subgrid prescriptions embedded in cosmological runs \citep{pfeffer2024origin, doppel2024imposters}. Yet, to our knowledge, no galaxy-scale simulations currently resolve individual GC formation in UDGs. This motivates our approach: we identify GCs self-consistently in high-resolution, galaxy-scale merger simulations that produce UDGs and analyze their internal properties, thereby moving beyond unresolved methods commonly applied to cosmological simulations.

In this work, we present a comprehensive model for the formation of UDGs and their GC populations via mergers of gas-rich dwarf galaxies. By employing high-resolution $N$-body and hydrodynamical simulations, we trace two merger simulations and study the properties of the remnant galaxies, which are found to be consistent with the UDG definition. Additionally, the remnants are found to host GC candidates (GCCs) whose properties we describe in detail. We aim to bridge theoretical predictions with observations and shed light on the diverse origins of UDGs.

\begin{figure*}
\centering
\includegraphics[width=\linewidth]{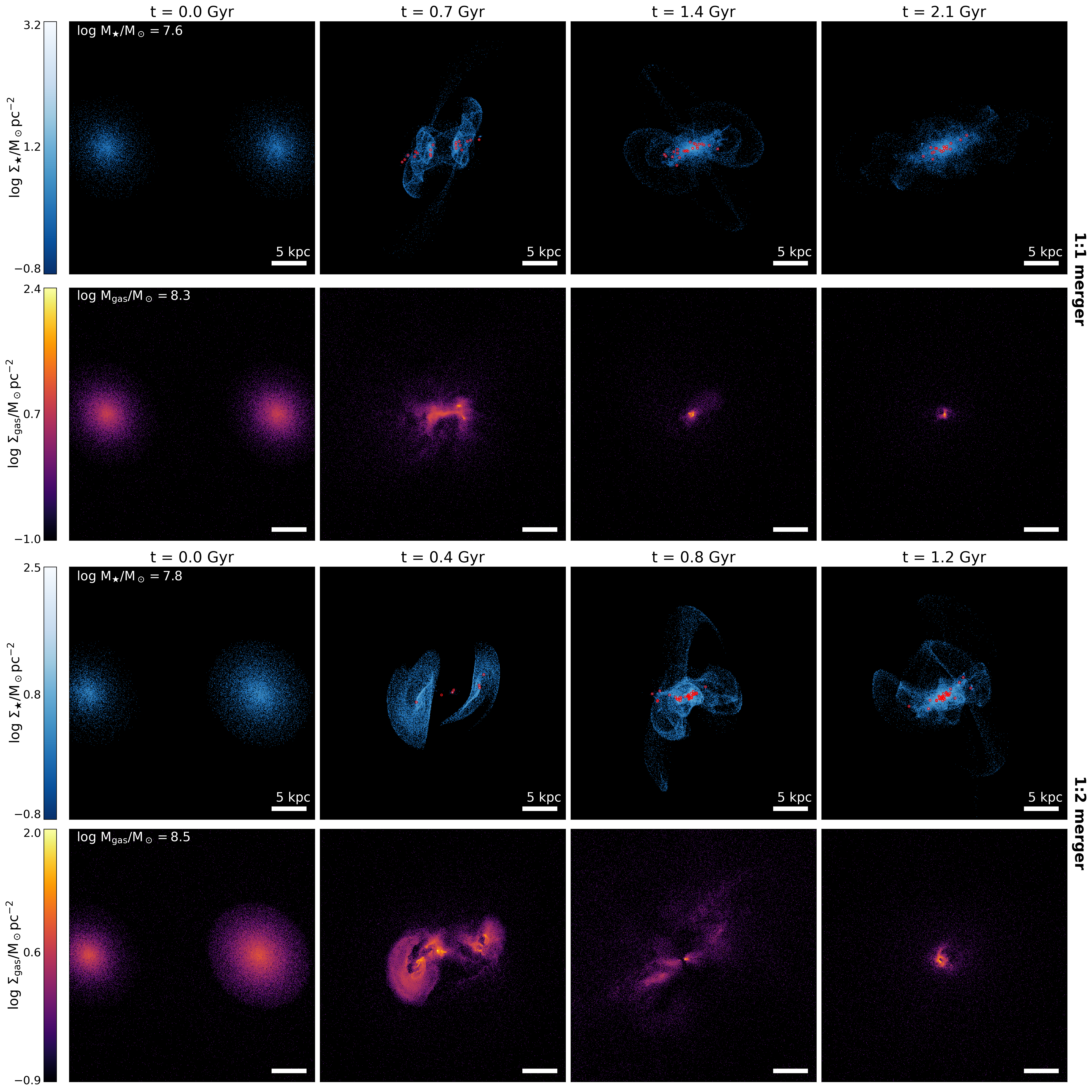}
\caption{Surface density snapshots along the z-axis for the merger simulations \textbf{A:A} (1:1) and \textbf{A:B} (1:2; galaxy properties in Table~\ref{tab:galaxies}). For each simulation, stellar (upper sub–row) and gas (lower sub–row) surface density maps are shown separately. The top two rows correspond to the 1:1 case and the bottom two to the 1:2 case. Columns show four epochs evenly spaced in time, from the initial conditions (left) to the final state (right). Red points mark the stellar overdensities identified with HDBSCAN in the stellar maps. The interactions disperse much of the gas from the remnant and trigger the formation of massive clusters. The post-merger galaxies show extended stellar spatial distributions consistent with UDGs.}
\label{fig:snapshots_mergers}
\end{figure*}

\section{Methods}

We use the Tree+GRAPE $N$-body/smoothed-particle hydrodynamics (SPH) code ASURA to study the formation of UDGs and the associated formation of massive SCs resulting from dwarf gas-rich mergers (see \citealt{saitoh2008toward, saitoh2009toward} for code details). Previously, we used ASURA to investigate massive SC formation in galaxy mergers \citep{2012ApJ...746...26M, matsui2019property, matsui2025formation}. Below, we summarize the simulation setup and star cluster detection pipeline.

\subsection{Simulations} 

\label{subsec:sims} % used for referring to this section from elsewhere

\begin{table}
    \centering
    \begin{tabular}{|c|c|c|}
        \hline
          & \textbf{A} & \textbf{B} \\
         \hline
        $M_\star$ [M$_\odot$] & $2\times10^7$ & $4\times10^7$\\
        \hline
        $M_{\rm gas}$ [M$_\odot$] & $10^8$ & $2\times10^8$\\
        \hline
        $M_{\rm halo}$ [M$_\odot$] & $5\times10^{10}$ & $10^{11}$\\
        \hline
        $h_D$ [kpc] & 1.5 & 2.1 \\
        \hline
        $r_h^{3D}$ [kpc] & 2.5 & 4.8\\
        \hline
        $r_s$ [kpc]& 7.9 & 10\\
        \hline
        $[\text{Fe/H}]_\text{gas}$ & -1.6 & -1.6\\
        \hline
    \end{tabular}
    \caption{Properties of the progenitor galaxies used to initialize the simulations. Rows list from top to bottom: stellar mass, gas mass, halo mass, disk scale length, stellar half-mass radius, NFW scale radius, and iron-to-hydrogen ratio in gas. We adopt $Z_\odot=0.014$.}
    \label{tab:galaxies}
\end{table}

ASURA treats dark matter (DM) and stars as collisionless $N$-body particles and gas with SPH particles. The particle masses are initially $2\times 10^3$ $\rm M_\odot$ which includes DM, gas, and stars. Star formation is permitted when a SPH gas particle satisfies T < 100 K, $n_\text{H} > 100$ $\rm cm^{-3}$, and $\nabla \cdot v < 0$; where T, $n_H$, and $v$ are the temperature, number density, and gas velocity, respectively. Newly formed stars have a mass equivalent to one third of their parent gas particle mass, namely $666.7$ $\rm M_\odot$. When a gas particle spawns a star particle, the gas particle loses mass equal to that of the star particle. We note that the median mass of newly formed stellar particles decreases to $\sim 500~\rm M_\odot$ due to stellar feedback.

Concerning the gas, the radiative cooling and photoionization heating rates from the UV background are computed using tabulated cooling/heating functions generated by CLOUDY \citep{2013RMxAA..49..137F}. These functions span
the temperature range $10$--$10^{9}$~K and depend on the gas temperature,
density and metallicity. We use a softening length of 1 pc for all particles in the simulation (i.e., DM, gas, and stars). The multiphase gas, the low mass of new star particles, and the small softening length allow us to resolve star cluster formation.

Each newly formed star particle resembles a stellar population that follows a Kroupa initial mass function \citep{kroupa2001variation}. Our simulations include thermal feedback and chemical enrichment, from Type II supernovae (SNe). Each SN deposits $10^{51}~{\rm erg}$ into the surrounding SPH particles. Metal enrichment is followed with Chemical Evolution Library (CELib) \citep{saitoh2017chemical}, including contributions from Type II and Type I SNe, AGB stars, and neutron-star mergers. We adopt IMF-averaged feedback, which should not differ significantly from IMF-sampled feedback given that our star particles spawn with masses of 666.7 $\rm M_\odot$, above the 500 $\rm M_\odot$ threshold below which \citet{10.1093/mnras/stab291} recommends IMF sampling.

We model two exponential gaseous+stellar discs embedded in NFW DM haloes \citep{navarro1996structure}. The initial distribution of particles is generated by MAGI \citep{2018MNRAS.475.2269M}. Galaxy \textbf{A} has disk scale length $h_D=1.5$ kpc, total stellar mass $M_\star=2\times 10^7$ $\rm M_\odot$, total gas mass $M_{\rm gas}=10^8$ $\rm M_\odot$, and total DM mass $M_\text{DM}=4.998\times10^{10}$ $\rm M_\odot$. The disk mass is $10^8$ $\rm M_\odot$: $80\%$ gas and $20\%$ stars. The halo has a mass of $5\times10^{10}$ $\rm M\odot$, including $0.04\%$ halo gas, and a scale radius $r_s=7.9$ kpc. Galaxy \textbf{B} follows the same proportions in disk and halo, but its mass is twice that of Galaxy A. It has a disk scale length of $h_{D}=2.1~{\rm kpc}$, total stellar mass of $M_{\star}=4\times 10^7~{\rm M_{\odot}}$, total gas mass of $M_{\rm gas}=2\times 10^8~{\rm M_{\odot}}$, and total DM mass of $M_{\rm DM}=9.996\times 10^{10}~{\rm M_{\odot}}$ with a scale radius of $r_s=10~{\rm kpc}$. The initial temperature of gas in the disk and halo is $10^{4}~{\rm K}$ and $10^{5}~{\rm K}$, respectively. The metallicity of gas starts at ${\rm [Fe/H]}=-1.6$, which is consistent with that observed in local dwarfs \citep{kirby2013universal}. The initial disk has the Toomre Q value of $1.5$. Table \ref{tab:galaxies} lists the parameters of the progenitor galaxies. We note that the progenitor galaxies have extended radii, which put them in the UDG regime (see Section \ref{sec:galaxies_properties}).

We perform three experiments: a 1:1 merger (\textbf{A}+\textbf{A}), a 1:2 merger (\textbf{A}+\textbf{B}), and an isolated evolution of galaxy \textbf{A}. The initial progenitor positions are $(x~[{\rm kpc}],y~[{\rm kpc}],z~[{\rm kpc}])=(-12.5,0,0)$ and $(12.5,0,0)$ for the 1:1 case and $(-16.7,0,0)$ and $(8.3,0,0)$ for the 1:2 case. The initial velocities are $(v_x, v_y, v_z) = (0, -4.6, 0)$ and $(0, 4.6, 0)$ km/s for the 1:1 run, and $(0,-7.8,0)$ and $(0,3.8,0)$ km/s for the 1:2 case. These choices place the galaxies on bound two-body orbits with the apocenter and pericenter distances of $25~{\rm kpc}$ and  $1~{\rm kpc}$, respectively. The disk spin axes of two galaxies are set to be $(0.35,0.35,0.87)$ and $(-0.35,-0.35,0.87)$ in all galaxy merger simulations. The isolated run evolves A alone to assess cluster formation in the absence of interactions. We note that no star cluster is placed in the initial conditions.

\subsection{Star cluster detection}

\label{subsec:sc_detection}

We identify bound stellar overdensities using HDBSCAN\footnote{\hyperlink{https://hdbscan.readthedocs.io/en/latest/index.html}{https://hdbscan.readthedocs.io/en/latest/index.html}} \citep{McInnes2017}. HDBSCAN (Hierarchical Density-Based Spatial Clustering of Applications with Noise) is a clustering algorithm designed to identify clusters of varying density and arbitrary shape within datasets, effectively distinguishing significant structures from noise, making it well suited to our simulated GC candidates. \citet{hunt2021improving} demonstrated that HDBSCAN is more sensitive and effective for identifying star clusters than DBSCAN or K-means. Additionally, the hyperparameter tuning in HDBSCAN is simpler than in the other clustering algorithms.

We cluster in 6D space ($x$, $y$, $z$, $v_x$, $v_y$, $v_z$), considering only new stellar particles (i.e., those which form from gas particles once the simulation starts). We checked that old stellar particles are not bound in any star cluster. Before we ran the clustering in the 6D space, we subtracted the median and scaled the data, setting the interquartile range to 1. We set the minimum cluster size hyperparameter to 100. Considering that the median new stellar particle mass is $\sim500~\rm M\odot$, the minimum mass of a cluster is $\sim5\times10^4$ $\rm M_\odot$, which aligns with the low mass end of GCs. We use the default value of the hyperparameter $\text{min$\_$samples}$, which equals the minimum cluster size. We checked that reasonable variations from these hyperparameter values do not change the number of star clusters detected or the characteristics of the clusters. 

For each cluster identified by HDBSCAN, we iteratively calculate the center of mass with shrinking spheres starting with the members within a 70 pc radius, with a convergence criterion of 0.1 pc. We initialize each iteration at the cluster medoid. Using this center of mass, we compute the 3D half-mass radius $r^{3D}_h$ considering particles within a 70 pc sphere. Subsequently, we calculate cluster properties using members within spherical apertures of radius $3~r_h^{3D}$. To focus on GC-like systems, we post-filter the star clusters by half-mass radius and mean density based on the Milky Way GC parameters reported in the \citet{baumgardt2018catalogue} catalog\footnote{We use the online updated catalog: \hyperlink{https://people.smp.uq.edu.au/HolgerBaumgardt/globular/}{https://people.smp.uq.edu.au/HolgerBaumgardt/globular/}}. We apply two filters: half-mass radius below that of the most extended MW GC ($r_h < 39.44$ pc), and mean density above that of the least dense GC ($\rho_h>0.036~\rm M_\odot/\rm{pc}^3$).

We inspect the surface density profiles of clusters obtained from the procedure above in the final simulation snapshots. The star clusters are consistent with single stellar clumps, as shown in Figures \ref{fig:gcs11} and \ref{fig:gcs12}. We find that the 1:1 merger simulation contains a putative nuclear star cluster (NSC) (see Section \ref{subsec:nsc}), which stands out from the rest of the cluster population. Excluding this cluster, we classify the remaining star clusters as globular cluster candidates (GCCs).

\section{Mergers and Isolated Evolution}

\begin{figure}
\centering
\includegraphics[width=1\linewidth]{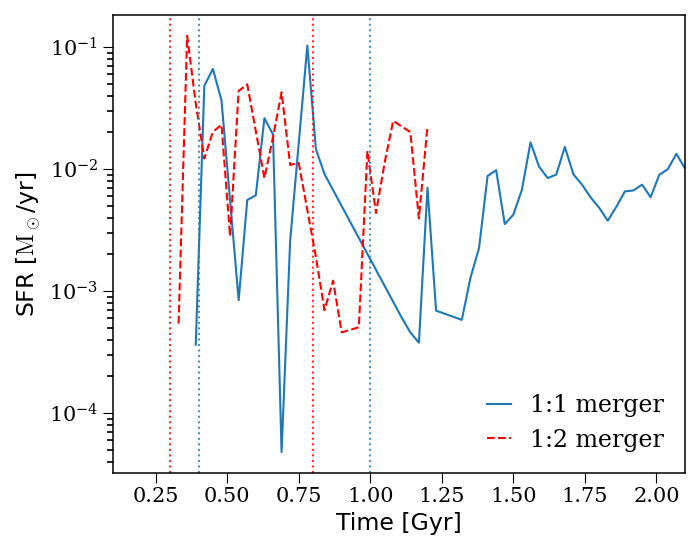}
\caption{Star-formation histories for the merger simulations \textbf{A:A} (1:1) and \textbf{A:B} (1:2). The 1:1 run is evolved to later times than the 1:2 run. Vertical dotted lines signal the times of first encounter and coalescence in each simulation following the color of the legend. Early peaks in the SFR coincide with pericentric passages, when tidal torques compress the gas and trigger starbursts. Around coalescence, both remnants enter a quiescent phase—shorter in the 1:2 case—before settling into sustained, secular star formation that persists to the end of the simulations (styles/colors as indicated in the legend).}
\label{fig:SFH_mergers}
\end{figure}

We present stellar and gas surface density snapshots of the merger simulations in Fig. \ref{fig:snapshots_mergers}. The left panels show the initial configurations and subsequent panels towards the right move forward in time with the rightmost panel showing the final state of the simulations. In both merger experiments, the galaxies experience two pericentric passages before coalescence; the remnant forms by $t\sim1$ Gyr in the 1:1 case and by $t\sim 0.8$ Gyr in the 1:2 case. The tidal shocks and orbit-driven compressions trigger the birth of compact stellar clusters in the central part of the remnants. We mark the positions of the stellar overdensities detected with red dots in Fig. \ref{fig:snapshots_mergers}. Consistent with this picture, the star-formation histories (SFHs) in Fig. \ref{fig:SFH_mergers} show initially sharp SFR spikes that coincide with the times of the encounters, followed by a quiescent period around coalescence. Finally, a secular star formation phase takes place after the remnants settle.

By contrast, the isolated dwarf galaxy (shown in Fig.~\ref{fig:snapshots_isolated}) evolves quasi-steadily, with modest redistribution of gas. The SFR shows a single, low peak of $\sim5\times10^{-4}$ M$_\odot/$yr at t = 1.2 Gyr. By the end of the simulation, at t = 1.4 Gyr, no GCCs are detected, and the galaxy retains its initial rotation, as expected for an unperturbed disc (see the right panel in Fig.~\ref{fig:galaxies_los}).

\section{Results}

\begin{table}
\centering
\caption{Properties of the post-merger and isolated galaxies at the end of the simulations considering 3D spatial information. The properties from top to bottom are the stellar mass ($M_\star$), gas mass ($M_{\text{gas}}$), new stellar mass ($M_{\star \text{new}}$), dynamical mass or total mass ($M_{\rm dyn}$), halo mass $M_\text{halo}$, 3D stellar half-mass radius ($r^{3D}_{h}$), stellar 1D velocity dispersion ($\sigma_{\star}$), number of GCCs ($N_{\text{GC}}$), total GCCs mass ($M_{\text{GC}}$), GCCs mass to stellar mass ratio ($M_{\rm GC}/M_\star$), half-mass radius of the system of GCCs ($r_{h,\text{GC}}$), half-number radius of the GCCs system ($r_{1/2,\text{GC}}$), GCCs 1D velocity dispersion ($\sigma_{\text{GC}}$), GCCs mean age ($\langle age\rangle_{\text{GC}}$), and GCCs mean iron to hydrogen ratio ($\left[\text{Fe/H}\right]_\text{GC}$). See the text for the details about the calculations of the properties. We adopt $Z_\odot=0.014$. The masses $M_\star$, $M_\text{gas}$, $M_{\star\rm new}$, $M_{\rm dyn}$, and the velocity dispersion $\sigma_\star$ are calculated considering particles within a sphere with radius $3~r_h^{3D}$ centered at the galactic center of mass.}
\label{tab:galaxy_properties}
\begin{tabular}{lccc}
\hline
\hline
Property & 1:1 merger & 1:2 merger & Isolated \\
\hline
$M_\star$ [$\rm M_\odot$] & 5$\times10^{7}$ & 7$\times10^{7}$ & 2$\times10^7$ \\
\hline
$M_{\text{gas}}$ [$\rm M_\odot$] & 1.2$\times10^{7}$ & 3.3$\times10^{7}$ & $7.6\times 10^7$ \\
\hline
$M_{\star \text{new}}$ [$\rm M_\odot$] & 1.6$\times10^7$ & 1.5$\times10^7$ & 2.3$\times10^4$ \\
\hline
$M_{\text{dyn}}$ [$\rm M_\odot$] & 1.4$\times10^9$ & 3.4$\times10^{9}$ & $10^9$\\
\hline
$M_{\rm halo}$ [$\rm M_\odot$] & $10^{11}$ & $1.5\times10^{11}$ & $5\times10^{10}$ \\
\hline
$r^{3D}_{h}$ [kpc] & 1.9 & 2.6 & 2.5 \\
\hline
$\sigma_{\star}$ [km/s] & 35.4 & 43.9 & 1.7 \\
\hline
$N_{\text{GC}}$ & 20 & 39 & 0 \\
\hline
$M_{\text{GC}}$ [$\rm M_\odot$] & 4.4$\times10^6$ & 9.8$\times10^6$ & -\\
\hline
$M_{\rm GC}/M_{\star}$ & 0.09 & 0.14 & -\\
\hline
$r_{h,\text{GC}}$ [kpc] & 1.3 & 1.1 & -\\
\hline
$r_{1/2,\text{GC}}$ [kpc] & 1.2 & 0.8 & -\\
\hline
$\sigma_{\text{GC}}$ [km/s] & 34.3 & 23.3 & - \\
\hline
$\langle age\rangle_{\text{GC}}$ [Gyr] & 1.5$\pm0.2$ & 0.7$\pm0.2$ & -\\
\hline
$\left[\text{Fe/H}\right]_\text{GC}$ & -1.3$\pm 0.1$ & -1.3$\pm 0.1$ & - \\
\hline
\end{tabular}
\end{table}

\subsection{Post-merger UDGs with massive star clusters}
\label{subsec:are_UDG?}

\begin{figure}
\centering
\includegraphics[width=1\linewidth]{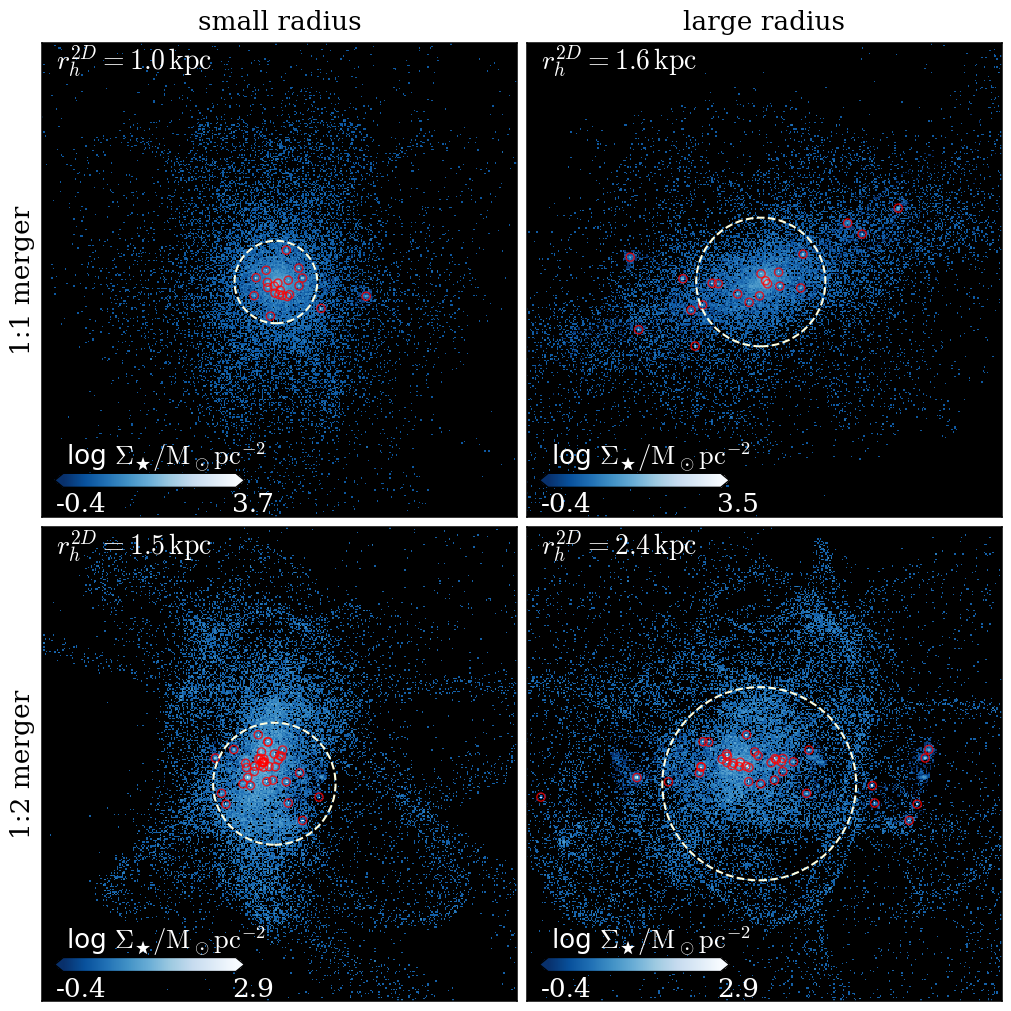}
\caption{Stellar surface–density maps of the merger remnants at the final snapshot. Top row: 1:1 remnant; bottom row: 1:2 remnant. We highlight the position of the cluster candidates with small red circles. For each remnant, two projections are shown: the left column is chosen to minimize the projected half-mass radius $r^{2D}_h$, and the right column to maximize it. Dashed big circles indicate $r^{2D}_h$ in each panel. In the 1:1 case, $r^{2D}_h\geq 1.5$ kpc for some viewing angles, whereas in the 1:2 case this condition holds for most viewing angles.}
\label{fig:post_merger}
\end{figure}

In Fig. \ref{fig:post_merger}, we show the stellar surface densities of the two post-merger galaxies at the end of the simulations under two projections, one that minimizes the radius (left column) and one that maximizes it (right column). We also highlight the positions of the GCCs detected as described in Section \ref{subsec:sc_detection}. We see that a system of GCCs is present in the remnants (we describe their properties in Section \ref{subsec:gcs_properties}). The galaxies resulting from our merger simulations exhibit disturbed characteristics with a triaxial morphology; hence, the projected radius of the galaxies depends on the viewing angle. We calculate the projected stellar half-mass radius $r^{2D}_h$ under random viewing angles for both galaxies (see Fig. \ref{fig:r2D_distribution}), finding that the 1:2 galaxy has $r^{2D}_h\gtrsim1.5$ kpc irrespective of the viewing angle, whereas this is true for $40\%$ of the viewing angles for the 1:1 remnant. The remnants have a mass of $5\times10^7~\rm$ and $7\times10^7~\rm M\odot$ for the 1:1 and 1:2 cases, respectively (see Table \ref{tab:galaxy_properties} for the full properties of the post-merger galaxies), which aligns with the values of low-mass UDGs and low-surface-brightness galaxies (LSBs) \citep{van2015forty, gannon2024catalogue, o2025wallaby}. Hence, the remnants follow the nature of UDGs being large and deficient of stellar mass. In Section \ref{subsec:observed_UDGs}, we show that the remnants satisfy the UDG definition based on the observational half-light radius and surface brightness.

\subsection{Properties of the galaxies}
\label{sec:galaxies_properties}

We present the main properties of galaxies at the end of the simulations in Table \ref{tab:galaxy_properties}. The masses involving baryonic matter and the velocity dispersions are calculated considering particles within a sphere with radius $3~r_h^{3D}$ centered at the galactic center of mass. This aperture encloses all detected GCCs and captures the bulk of the bound baryons.

The stellar half-mass radii of the remnants are $r^{3D}_h=1.9$ kpc (1:1) and 2.6 kpc (1:2). Such radii are larger than the mean half-mass radii of dwarf galaxies in the TNG50 cosmological simulations \citep{celiz2025mass}. In the longer run (1:1), $r^{3D}_h\sim2$ kpc for the last 0.5 Gyr, indicating that the extended structure is not a transient feature. In the 1:2 run, the radius holds a high value for 0.4 Gyr.  We include a time-series panel of $r^{3D}_h$ in the Appendix and reference it here (see Fig. \ref{fig:radius_evolution}).

\begin{figure*}
\centering
\includegraphics[width=1\linewidth]{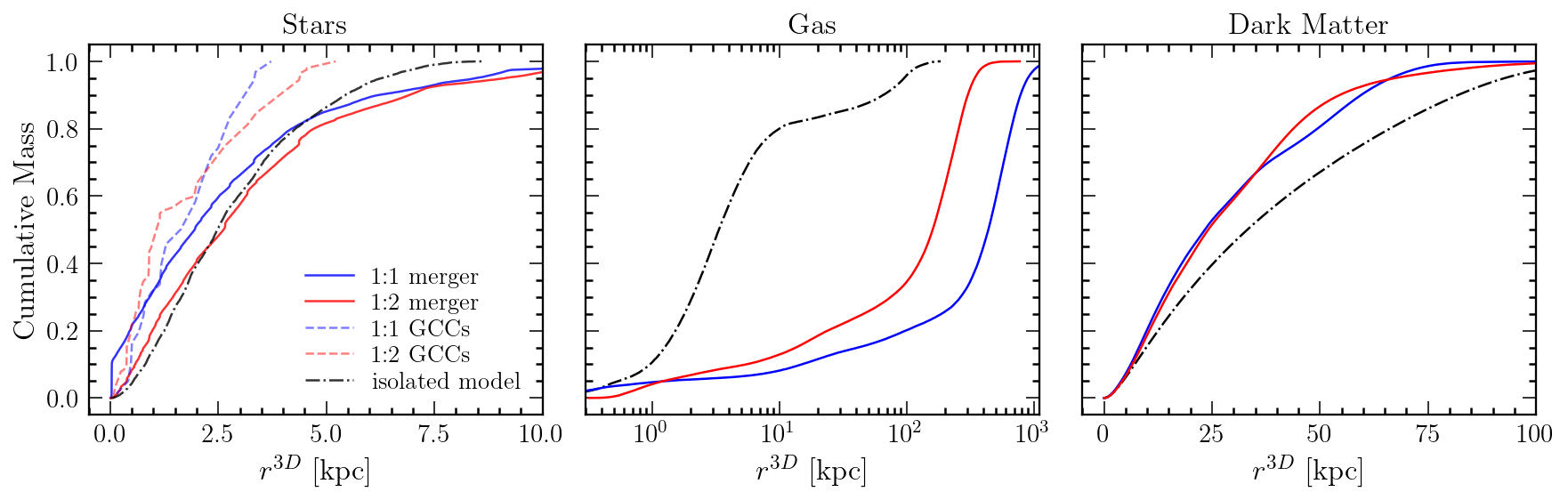}
\caption{Normalized three-dimensional mass profiles of stars, gas, and dark matter at the final snapshot of the three simulations used in this work (styles/colors as indicated in the legend). The isolated galaxy corresponds to the \textbf{A} (or 1) progenitor of the merger runs. All systems exhibit extended stellar profiles for their dwarf stellar masses. Next to each stellar profile, we show the corresponding radial profile of the globular-cluster candidates (GCCs) in dashed lines; the GCC system is more centrally concentrated than the field stars. In the post-merger galaxies, a large fraction of the gas is dispersed to radii of order $\gtrsim100$ kpc and remains only weakly bound. The DM haloes of the post-mergers display similar shapes and extend to $\sim100$ kpc.}
\label{fig:profiles}
\end{figure*}

\begin{figure}
\centering
\includegraphics[width=1\linewidth]{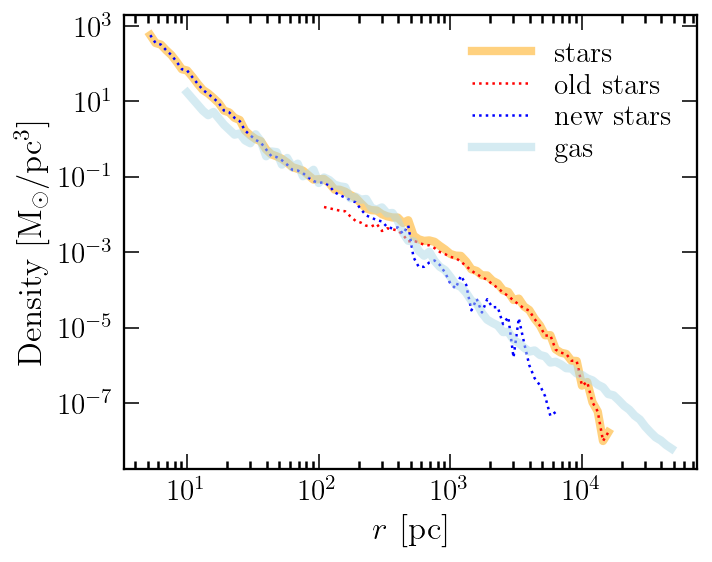}
\caption{Density profiles of stars and gas in the remnant of the 1:1 merger at the end of the simulation. The profiles probe the galaxy from its center, placed at the NSC center of mass, out to its edge beyond 10 kpc, with the stellar density spanning more than ten orders of magnitude. In the central region, new stellar particles dominate the stellar profile, and the high gas density there suggests ongoing accretion and continued in-situ star formation in the NSC. In the outskirts, the stars inherited from the progenitors dominate. Note that the profile extends only down to 5~pc and therefore does not resolve the denser, innermost region of the NSC.}
\label{fig:density_profile_stars11}
\end{figure}

We show the cumulative mass profiles of stars, gas, and dark matter for each galaxy by the end of the simulations in Fig.~\ref{fig:profiles}, with the total masses tabulated in Table~\ref{tab:galaxy_properties}. Additionally, we show stellar and gas density profiles of the more evolved 1:1 remnant in Fig.~\ref{fig:density_profile_stars11}. Concerning the stellar mass distribution, the merger remnants show more extended stellar profiles than the isolated control simulation, given that the mergers scatter a small portion of stellar particles to large radii. The density profile of the 1:1 remnant shows that within $\sim0.5$~kpc the new stars formed during the merger dominate the profile, whereas the old stellar population of the progenitors dominates beyond 0.5~kpc. The near-vertical increase in the cumulative 1:1 stellar profile and the high stellar density in the density profile at the center signal the presence of an NSC (see Section~\ref{subsec:nsc}). The stellar density rises to $\sim10^{3}~\mathrm{M_{\odot}\,pc^{-3}}$, comparable to the densities of the massive GCs in the MW (grey points in Fig.~\ref{fig:struct_gcs_props}).

In Fig.~\ref{fig:density_profile_stars11}, the young stellar profile of the 1:1 remnant closely traces the gas, except in the outskirts and at the very center. In the outskirts, the gas is too diffuse to form stars, so it is present without an accompanying young stellar component. At the center, conversely, the young stars greatly outnumber the gas: they represent the cumulative product of star formation, whereas the gas traces only the remaining reservoir, which has been depleted by star formation and feedback. The negative gradient in the gas profile suggests gas inflow towards the center of the galaxy. The gas reservoir peaks at
$\sim10~\mathrm{M_{\odot}\,pc^{-3}}$ ($n_{\mathrm{H}}\sim300~\mathrm{cm^{-3}}$), within the range of giant molecular cloud densities and above the star-formation threshold adopted in our
simulations. Hence the center of the galaxy can host star formation and increase the NSC mass. To confirm such star formation activity, the star formation history of the NSC in Fig.~\ref{fig:NSC_sfh} shows active star formation by the end of the 1:1 simulation.

The progenitors are gas dominated $M_{\rm gas}/M_{\star}=5$ and the isolated simulation remains gas dominated by the end of the simulation. However, the post-merger galaxies are stellar dominated with $M_{\rm gas}/M_{\star}=0.24$ and 0.47 in the 1:1 and 1:2 cases, respectively. This occurs because the intense star formation that takes place during the encounters leads to a large number of SNe II over a short period of time, expelling the gas from the galaxy. The gas does not fall back into the galaxy by the end of the simulations; hence, the cumulative mass profiles in Fig.~\ref{fig:profiles} show that the gas particles in the remnants lie well beyond the DM halo of the galaxy. 

\begin{figure*}
    \centering
    \includegraphics[width=\linewidth]{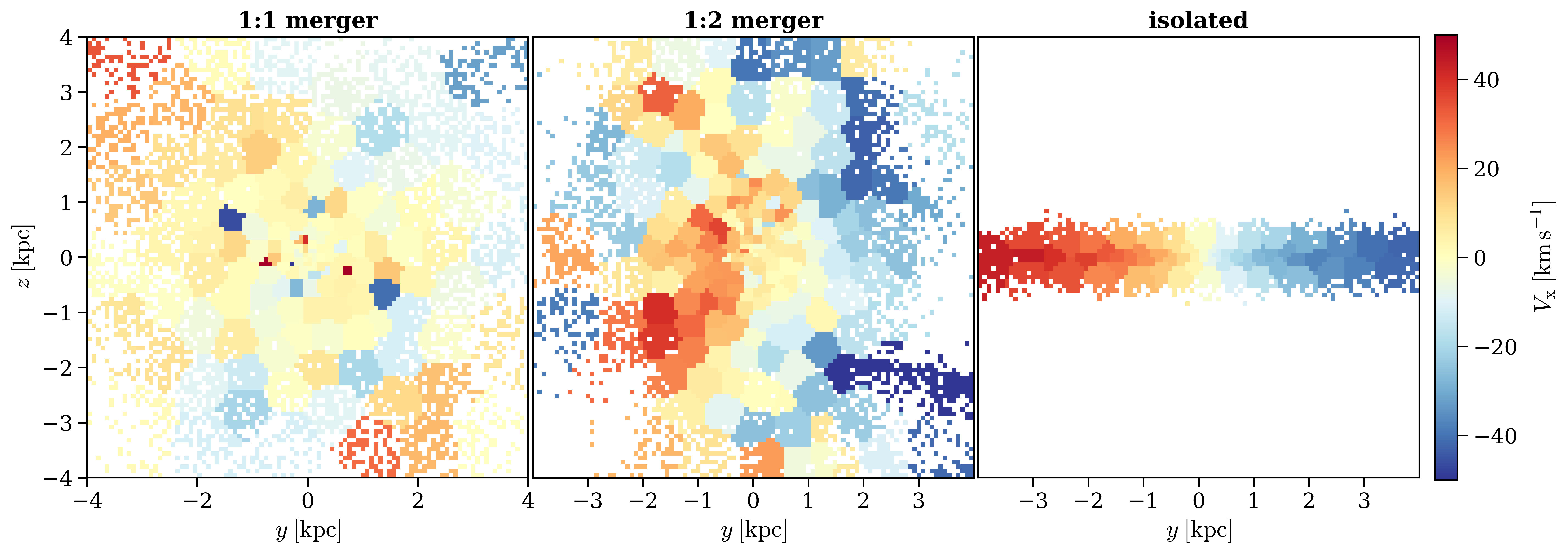}
    \caption{Line-of-sight stellar velocity maps ($V_x$) of the three galaxies at the end of the simulations: the 1:1 merger remnant (left), the 1:2 merger remnant (middle), and the isolated model (right). For each system, the $z$-axis is aligned with the galaxy's total stellar angular-momentum vector, so the maps show the velocity field viewed edge-on with respect to the net spin, i.e.\ the orientation that maximizes the projected rotation signal; the bulk (systemic) velocity of each galaxy has been subtracted. The GCCs have been removed from the maps. The 1:1 remnant shows no clear signal of rotation, whereas the 1:2 shows a large scale velocity gradient lumpy and asymmetrical. By contrast, the isolated galaxy retains the rotational pattern imprinted by the initial conditions. The maps adopt a 2D Voronoi binning using the PowerBin method by \citet{Cappellari2025}.}
    \label{fig:galaxies_los}
\end{figure*}

Concerning the dark-matter content, each remnant is embedded in a single, approximately spherical halo. Its smooth, relaxed morphology indicates that the two progenitor haloes have dynamically virialized into a common structure, and its total mass is consistent with the sum of the two progenitors, confirming that the dark matter is conserved during the merger. The remnant haloes are more centrally concentrated than the isolated model (see the right panel in Fig.~\ref{fig:profiles}) and extend out to $\sim100$~kpc from the center of mass.

The progenitors have NFW cuspy profiles, and we test whether the remnants show a cusp or core profile at the center. We calculate the density profiles (Fig.~\ref{fig:dm_density_profiles}) and derive their logarithmic slopes. Inside $\sim200$~pc the enclosed particle number becomes small enough that two-body relaxation can artificially flatten the profile and produce a spurious core \citep{2003MNRAS.338...14P}; we therefore restrict our interpretation to $r>200$~pc. Across this converged region both remnants and the isolated galaxy retain a negative logarithmic slope that steepens with increasing radius, with no flattening towards zero. Hence, the merger does not change the cuspy profiles of the progenitors into a core in the remnants.

Fig.~\ref{fig:galaxies_los} shows the line-of-sight velocity maps of the two remnants and the isolated model, each rotated into the frame in which the line of sight maximizes the projected rotation signal. Both merger remnants are dispersion-dominated, but they differ in their residual ordered motion. The 1:1 remnant (left) shows a patchy, low-amplitude field ($|V_x| \lesssim 20\,\mathrm{km\,s^{-1}}$) with no systematic large-scale gradient, indicating little coherent rotation. The 1:2 remnant (middle), by contrast, displays a weak but visually apparent large-scale gradient, with positive velocities on the $y<0$ side and a coherent negative-velocity region reaching $|V_x| \sim 40\,\mathrm{km\,s^{-1}}$ on the $y>0$ side. This gradient runs in the same sense as ordered rotation, suggesting that the 1:2 merger retains part of the main progenitor initial spin, in contrast to the 1:1 remnant where it is more thoroughly erased. The field nonetheless remains spatially lumpy and asymmetric rather than forming a clean disc-like dipole.

\begin{figure}
\centering
\includegraphics[width=1\linewidth]{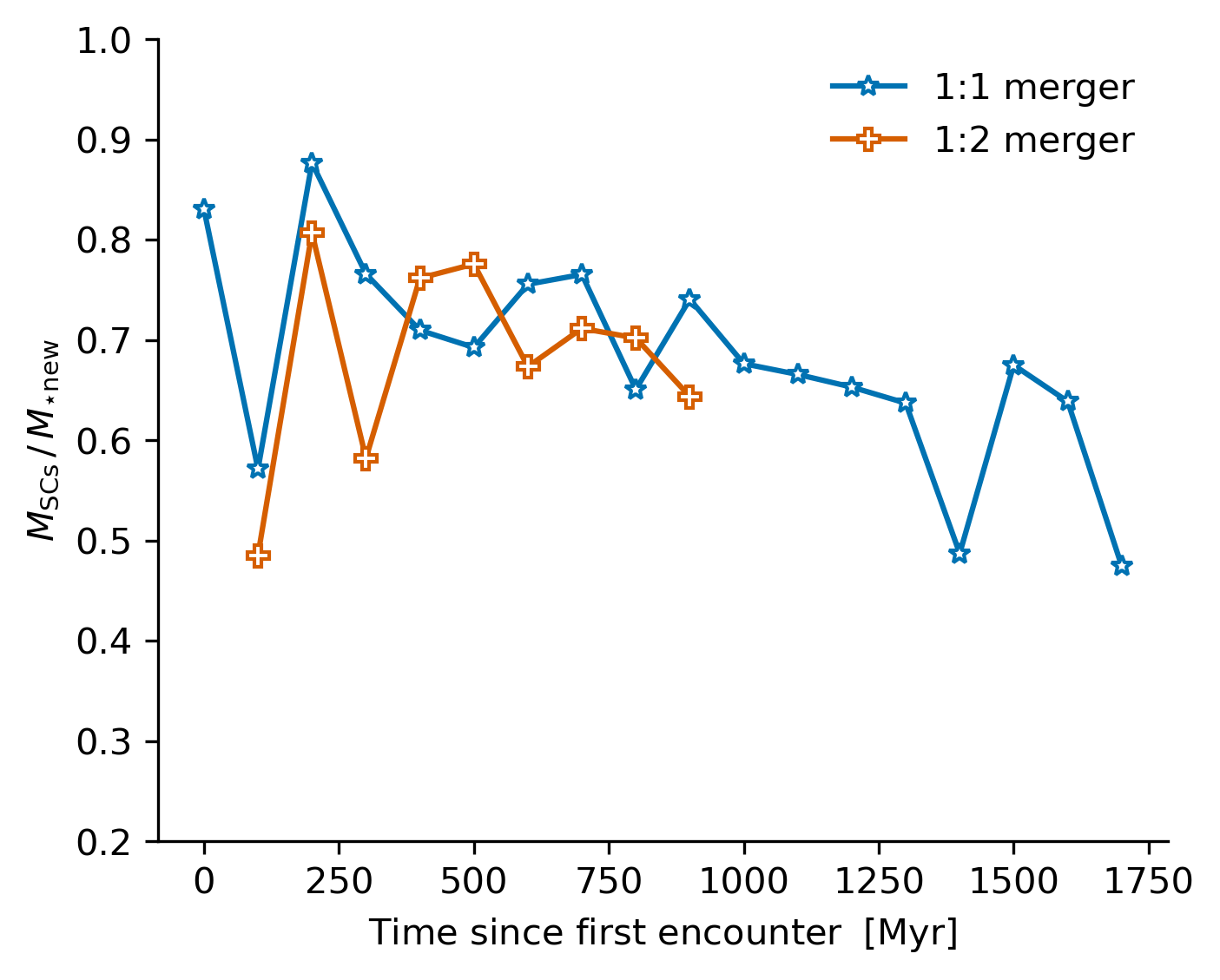}
\caption{Evolution of the total stellar mass in massive star clusters per total stellar mass formed in the two merger simulations. Throughout the runs, this fraction remains high ( $\gtrsim0.5$), indicating that star formation proceeds predominantly in compact, GC-like overdensities that persist over time. The time of the first encounter in each simulation is shown in Fig.~\ref{fig:SFH_mergers}.}
\label{fig:scs_fraction}
\end{figure}

Using the reference frame of Fig.~\ref{fig:galaxies_los} (i.e.\ aligned angular momentum, bulk velocity removed, and GCCs masked), we decompose the velocity of each stellar particle into cylindrical coordinates and adopt the azimuthal component $v_\phi$ as its rotational velocity. We then bin the particles in cylindrical radius and take the mean $v_\phi$ in each bin as the bulk rotation curve. After subtracting this mean rotation from each particle, we compute the 1D stellar velocity dispersion $\sigma_\star$, reported in Table~\ref{tab:galaxy_properties}; for the 1:1 and 1:2 remnants $\sigma_\star = 35.4$ and $43.9\,\mathrm{km\,s^{-1}}$, respectively. Applying the virial estimator $M = 3\,r\,\sigma_\star^2/2G$ using the half-mass radius of each remnant, we find that the virial masses agree with the dynamical masses $M_\mathrm{dyn}$ (or total masses) reported in Table~\ref{tab:galaxy_properties} to within a factor of $\sim2$, consistent with the remnants being near virial equilibrium. Finally, the ratio of the mass-weighted mean rotational velocity to $\sigma_\star$ is $V/\sigma = 0.06$ for the 1:1 remnant and $0.32$ for the 1:2, confirming that both systems are dispersion-supported.

Fig.~\ref{fig:scs_fraction} shows the time evolution of the mass locked in massive SCs per total stellar mass formed, $M_{\rm SCs}/M_{\star\text{new}}$, where $M_{\rm SCs}$ is the total mass of the SCs detected by our pipeline as described in Section \ref{subsec:sc_detection}. We remind the reader that these massive SCs have masses comparable to GCs, and the only distinction from the GCCs is that the latter were visually inspected to be consistent with single stellar clumps and different from an NSC. During the merger-driven bursts, the majority of star formation is clustered, leading to \(M_{\rm SCs}/M_{\star\text{new}}\gtrsim 0.5\) right after the first pericenter passage. This high fraction holds during the simulations as the SFR is low after the encounters (see Fig. \ref{fig:SFH_mergers}). We also report in Table~\ref{tab:galaxy_properties} the total GCCs mass to total stellar mass ratio $M_{\rm GC}/M_{\star}$ at the end of the simulations, $9\%$ and $14\%$ for the 1:1 merger and 1:2 merger, respectively.

\subsection{Properties of the GC candidates}
\label{subsec:gcs_properties}

We detect 20 and 39 GCCs in the 1:1 and 1:2 runs, respectively, using the pipeline described in Section \ref{subsec:sc_detection}. Surface density maps and catalogs with the properties of individual clusters are given in Appendix \ref{app:scs_supplementary}. Most GCCs appear as compact, centrally concentrated stellar clumps, with surface density rising toward their centers. Additionally, we report the properties of the GCC systems of the remnants in Table \ref{tab:galaxy_properties}. Below, we summarize the main findings.

\begin{figure}
\centering
\includegraphics[width=1\linewidth]{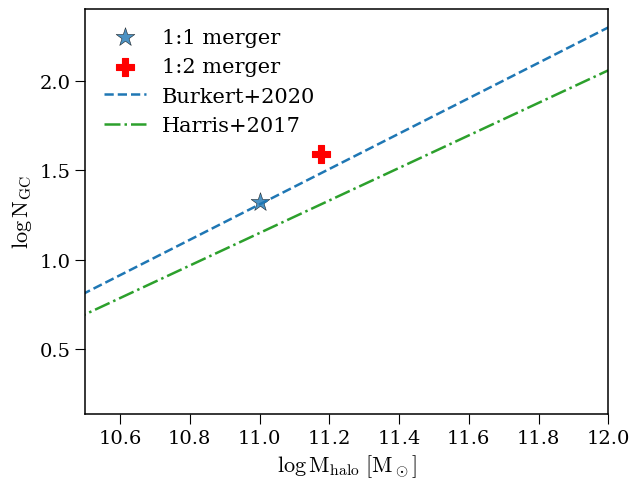}
\caption{Number of GCs versus halo mass of the merger simulations. Our results agree with the relation derived by \citet{burkert2020high}, while lying above the \citet{harris2017galactic} relation. Although the merger progenitors do not contain any SCs, the post-merger galaxies exhibit a rich population of massive SCs.}
\label{fig:gnhr}
\end{figure}

In the \(1{:}2\) remnant, the GCC system is roughly twice as massive and twice as numerous as in the \(1{:}1\) remnant (see Table~\ref{tab:galaxy_properties}); however, when the mass bound in the NSC is added to the \(1{:}1\) budget, the total mass in clusters becomes comparable between the two remnants, \(\sim 10^{7}\,\rm M_\odot\). The mean ages of the cluster systems are consistent with formation during the galactic encounters. In both remnants, the GCCs are more centrally concentrated than the field stars, with smaller half–mass/half–number radii ($r_{h,\rm GC} < r_h$) and steeper cumulative mass profiles (see the left panel of Fig.~\ref{fig:profiles}).

\begin{figure}
\centering
\includegraphics[width=0.9\linewidth]{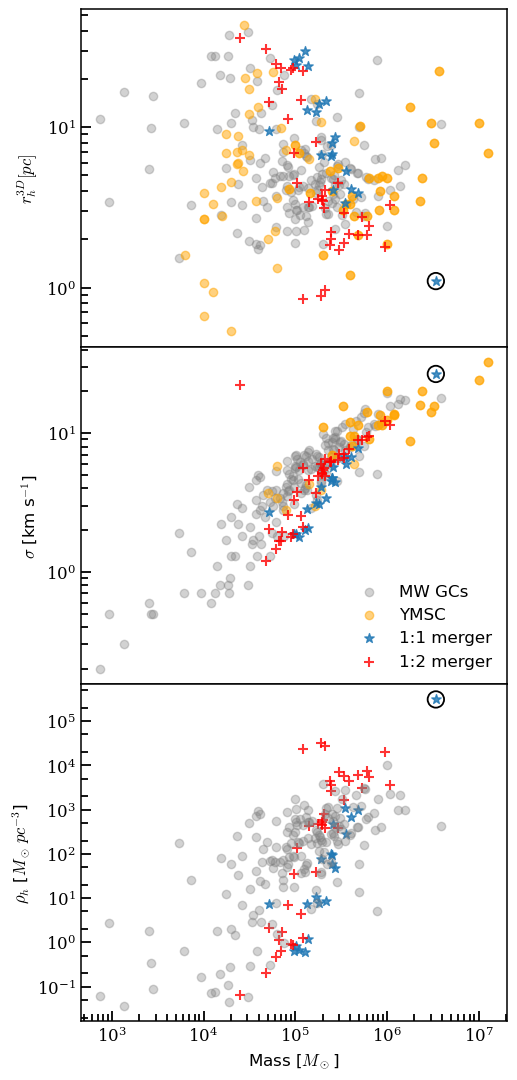}
\caption{Structural scaling relations of the globular-cluster candidates (GCCs) in the post-merger remnants. Panels show the 3D half-mass radius $r^{3D}_h$, 1D velocity dispersion $\sigma$, and half-mass density $\rho_h$ versus stellar mass of the clusters. For reference, we overlay Milky Way GCs from \citet{baumgardt2018catalogue} and young massive clusters compiled by \citet{annurev:/content/journals/10.1146/annurev-astro-081309-130834} and references therein. The nuclear star cluster (NSC) formed in the 1:1 simulation is signaled by the black circle. The simulated GCCs follow the observed sequences and largely overlap their loci, indicating broadly similar global dynamics.}
\label{fig:struct_gcs_props}
\end{figure}

The number of GCCs detected at the end of the simulations, 20 (1:1 case) and 39 (1:2 case), agrees with the GC number-halo mass relation (GNHR) \citep{burkert2020high, harris2017galactic}, as seen in Fig.~\ref{fig:gnhr}. We note that, observationally, the number of GCs in a galaxy is estimated by doubling the number of clusters brighter than the turnover point in the GC luminosity function, whereas in our simulations we can resolve and directly count the low-mass GCCs. The mass distributions of the two GCC systems are symmetric about approximately the same peak mass value (see Fig.~\ref{fig:gcmf}). Hence, doubling the number of GCCs more massive than the turnover mass of $2\times10^5~\rm M_\odot$ yields 20 (1:1 case) and 36 (1:2 case), in agreement with the fully resolved counts. Observationally, GC-rich UDGs are found to follow the GNHR
\citep{forbes2024ultra, forbes2025some}.

We report the velocity dispersion of the GCC systems, $\sigma_\text{GC}$, in Table~\ref{tab:galaxy_properties}. The values of $\sigma_\text{GC}$ and $\sigma_\star$ agree more closely in the 1:1 remnant than in the 1:2 case. We regard the agreement $\sigma_\text{GC} \approx \sigma_\star$ in the 1:1 remnant as the more reliable result, since this remnant has had more time to settle, whereas the 1:2 remnant shows disturbed features by the end of the simulation.

Fig.~\ref{fig:struct_gcs_props} contrasts the structural properties of the simulated GCCs with observed Milky Way GCs and young massive star clusters (YMSCs). The GCCs populate a similar locus in the mass-$r_{h}^{3\mathrm{D}}$ and mass-$\rho_h$ planes as the observations. The mass-$\sigma$ relation is tight, and the GCCs fall on the same sequence as the observations. We note that our dispersion values are lower limits due to the softening length used in the simulations. The circled symbol marks the nuclear star–cluster candidate in the \(1{:}1\) remnant; it extends these trends to \(M\sim 10^{7}\,\rm M_\odot\) with \(\sigma\) and \(\rho_h\) consistent with an extrapolation. Overall, the agreement across \(r_h\), \(\sigma\), \(\rho_h\), and mass argues that the simulated clusters reproduce the global structural properties of real star clusters, despite finite spatial and mass resolution that may limit the internal structure.

\begin{figure}
\centering
\includegraphics[width=1\linewidth]{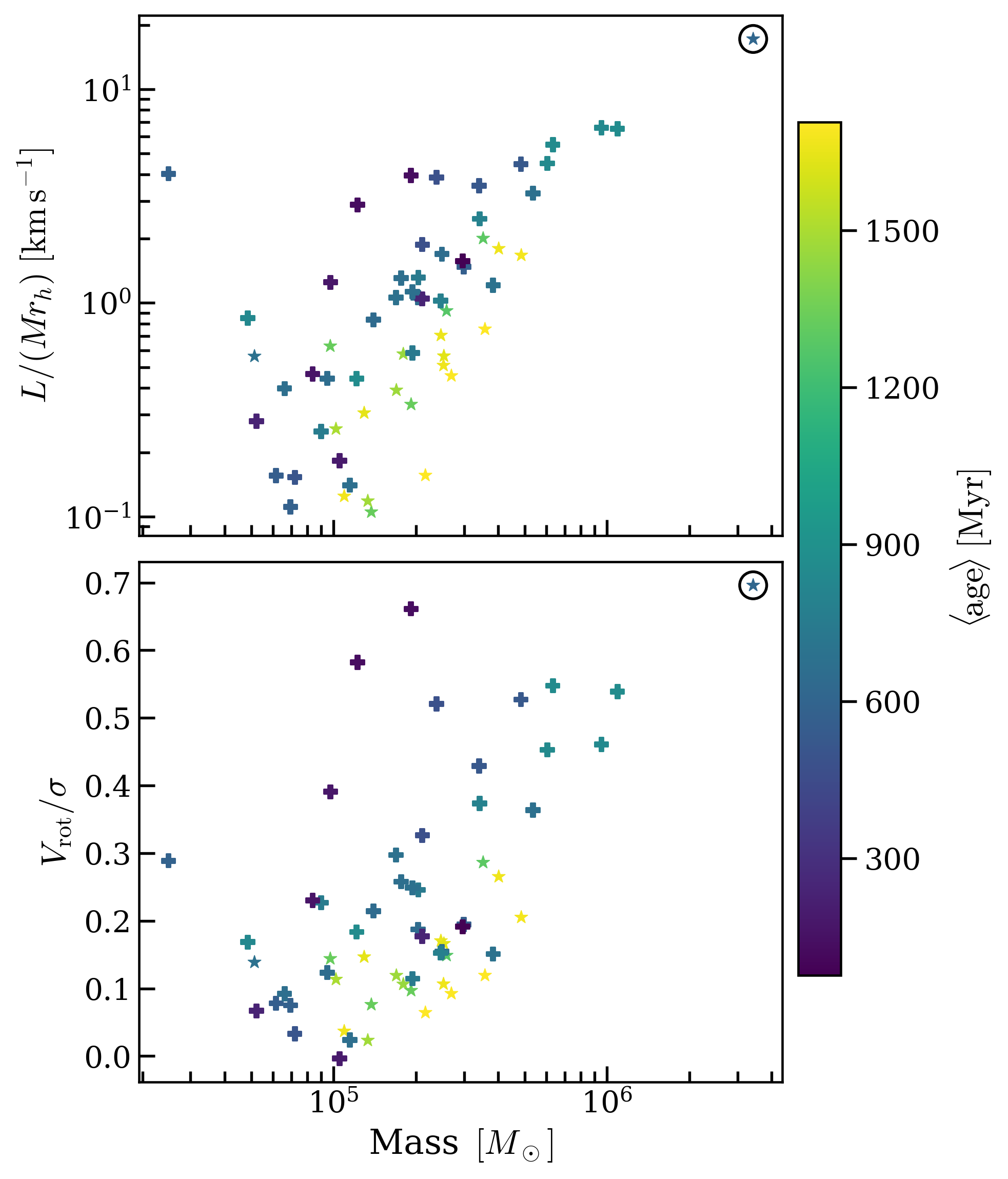}
\caption{Internal rotation of the GCCs in the post-merger remnants. Top panel: rotation proxy—specific angular momentum per radius—versus stellar mass. Bottom panel: mean rotation-to-dispersion ratio $V_\text{rot}/\sigma$ versus stellar mass. Star symbols denote GCCs from the 1:1 run; cross symbols denote the 1:2 run. The NSC in the 1:1 case is highlighted with a black circle. Point color encodes cluster age. More massive clusters rotate faster and exhibit larger $V_\text{rot}/\sigma$, and rotation strength declines with age.}
\label{fig:rotation_v_mass}
\end{figure}

We quantify cluster rotation by first selecting particle members within $3\,r^{3D}_h$ from the center of mass. For each cluster, we compute the total angular momentum $\mathbf{L}$ and align the $z$-axis with it. We then estimate the rotation velocity $V_\text{rot}$ of a cluster as the mean tangential velocity in cylindrical coordinates. As a proxy for rotational speed, we also use the specific angular momentum divided by the half-mass radius, $L/(M r_h)$. The upper panel of Fig.~\ref{fig:rotation_v_mass} shows $L/(M r_h)$ versus cluster mass for the two merger simulations. More massive clusters rotate progressively faster. At fixed mass, younger clusters rotate faster, consistent with the systematic offset towards higher rotation in the 1:2 merger relative to the 1:1 merger, which reflects the younger ages of the former. This trend is naturally explained by dynamical evolution: two-body relaxation progressively erodes ordered rotation.

The lower panel of Fig.~\ref{fig:rotation_v_mass} presents the ratio $V_{\rm rot}/\sigma$. Despite our clusters being dispersion supported ($V_{\rm rot}/\sigma<1$), the contribution from ordered rotation increases with mass. In Appendix Fig.~\ref{fig:gcs_los_velocity} we highlight the rotation of GCCs by showing the line-of-sight velocity maps; the strongest rotators show point-antisymmetric maps about the origin, with the approaching and receding sides producing a clear velocity gradient across the center. The correlation of rotation (e.g.\ tangential velocity or angular momentum) with cluster mass has been reported observationally in MW GCs \citep{bianchini2018internal} and in simulations of massive clusters formed in dwarf galaxy mergers \citep{2020ApJ...904...71L}, supporting our findings. The origin of the rotation in our clusters is beyond the scope of this work. It has been shown that gas clouds acquire angular momentum from tidal torques and that star clusters inherit this rotation from their natal clouds \citep{2020ApJ...904...71L, 2017MNRAS.467.3255M}.

\begin{figure}
\centering
\includegraphics[width=1\linewidth]{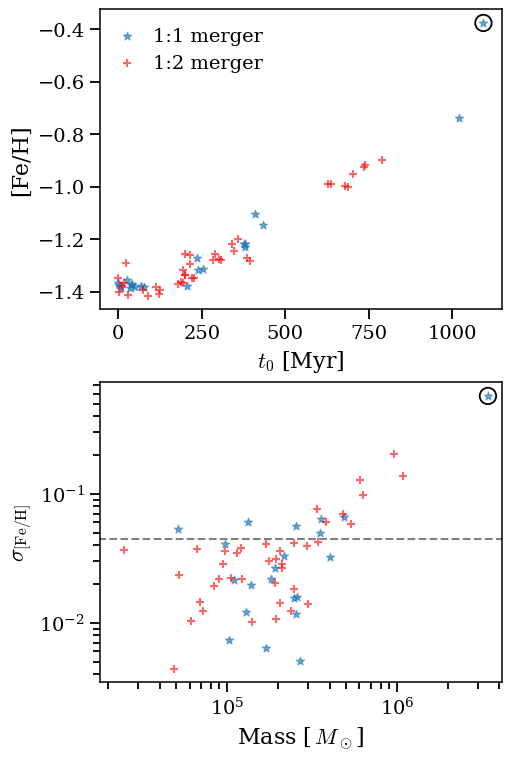}
\caption{Top panel: mean iron abundance [Fe/H] versus formation time, with time shifted so that the first-formed cluster occurs at $t_0=0$. Metallicity increases over time as the earliest clusters enrich the surrounding interstellar medium. Bottom panel: metallicity spread versus cluster stellar mass; higher-mass clusters tend to exhibit larger metallicity dispersions (styles/colors as indicated in the legend). The dashed line marks the median metallicity spread of 0.045 in MW GCs, as reported by \citet{Bailin_2019}.}
\label{fig:metallicity_corr}
\end{figure}

We quantify cluster metallicities by computing, for each GCC, the mean [Fe/H] and its dispersion $\sigma_\text{[Fe/H]}$ of stellar members within $3r_h^{3D}$. The upper panel of Fig.~\ref{fig:metallicity_corr} shows the metallicity of clusters versus the formation epoch $t_0$ (scaled by the formation of the first GCC in the simulation). Both merger simulations display a clear enrichment sequence: the earliest clusters form from metal--poor gas, while subsequent generations become progressively more metal-rich as stellar feedback enriches the surrounding interstellar medium. This trend resembles the age--metallicity relation observed in galaxies and their star cluster systems \citep{2022A&A...666A..80N, 1998MNRAS.296.1045C}. The lower panel of Fig.~\ref{fig:metallicity_corr} presents the metallicity dispersion as a function of the cluster mass. We find a positive correlation, indicating that more massive clusters exhibit broader metallicity spreads. This is consistent with the catalogue of intrinsic iron spreads in Milky Way GCs of \citet{Bailin_2019}, who find that the most luminous clusters ($L \gtrsim 10^5\,L_\odot$) show progressively larger iron dispersions, while most clusters have small but non-zero spreads of $\sigma_0 < 0.1$~dex with a median of $0.045$~dex; we recover a median of $0.029$~dex. We note, however, that our values likely represent lower limits, since the finite mass resolution of our simulations (each stellar particle carries a single abundance) does not resolve abundance variations on sub-particle scales. Theoretical studies attribute the origin of spread metallicity to the merging of chemically distinct gas clumps together with self-enrichment by SN ejecta retained within the forming cluster \citep{2021MNRAS.507..834M, 2018ApJ...863...99B}.

\subsection{Nuclear star cluster}
\label{subsec:nsc}

We find a nuclear star cluster (NSC) in the post-merger UDG of the 1:1 merger simulation. Its high mass causes a near-vertical increase in the stellar profile around the center of the galaxy (see Fig. \ref{fig:profiles}) as well as a peak in the density profile consistent with massive star clusters (see Fig.~\ref{fig:density_profile_stars11}). It contains $\sim10\%$ of the total stellar mass. In Fig. \ref{fig:struct_gcs_props}, we see that this cluster, signaled by a black circumference, is more massive than the other SCs. It is also compact, showing the highest density and a small half-mass radius. The GCCs in the 1:1 merger have an age spread of 70 Myr at most with a median age spread of $\sim 2$ Myr, whereas the NSC has an age spread of $\sim 500$ Myr. Such age spread is also related to its metallicity spread, which is the highest among the massive SCs (see Fig. \ref{fig:metallicity_corr} lower panel). We suggest that nuclei of UDGs formed by mergers could have such wide spreads in ages and metallicities.

\begin{figure}
\centering
\includegraphics[width=1\linewidth]{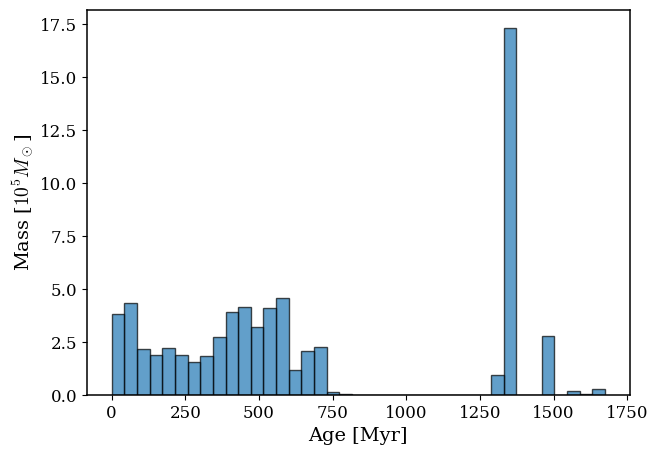}
\caption{Age distribution of the stellar particles that constitute the nuclear star cluster (NSC) in the 1:1 post-merger remnant. Unlike the GCCs—which are dominated by a single star-formation burst—the NSC exhibits a composite history: two approximately coeval episodes (associated with the assembly of GCCs) together with sustained, in-situ star formation lasting $\sim750$ Myr until the end of the simulation.}
\label{fig:NSC_sfh}
\end{figure}

We show the formation history of the NSC of the 1:1 merger in Fig.~\ref{fig:NSC_sfh}. The oldest star particles in the NSC formed in the two initial starbursts; we confirm that they were members of massive star clusters in early snapshots and were not yet located at the center of the galaxy. For the following 500 Myr no further members formed, coinciding with the quenching of the galaxy after coalescence (see Fig.~\ref{fig:SFH_mergers}). After this inactive phase, members formed continuously until the end of the simulation. By then, the gas density at the center of the galaxy is high enough to sustain star formation (see Fig.~\ref{fig:density_profile_stars11}), and the continuous late-time part of the NSC's SFH suggests that this central activity has been ongoing for several hundred million years. The NSC in our simulation is assembled through two channels: the ex situ infall and merging of GCs towards the galactic center, potentially driven by dynamical friction, together with continuous in situ star formation from the central gas. This formation route is consistent with our previous results, in which star and gas clumps aggregate at the galactic center either through direct merger or tidal disruption, both in dwarf mergers \citep{matsui2025formation} and in isolated gas-rich low-mass disc galaxies \citep{bekki2007formation}. These two channels also appear in other simulations of dwarf galaxies: \citet{2025OJAp....8E.146G} show that an NSC can grow ex situ through the merging of SCs within a dwarf galaxy, whereas \citet{2025MNRAS.539.1167G} find that a dwarf--dwarf merger drives gas into a central reservoir that builds an NSC in situ. Both channels have also been identified observationally by \citet{2024A&A...687A..83F} and \citet{2022A&A...667A.101F}.

\section{Discussion}
\label{sec:discussion}

\subsection{Do the remnants satisfy observational UDG criteria?}
\label{subsec:observed_UDGs}

The idealized nature of our simulations prevents a direct calculation of the half-light radius $r_e$ and surface brightness $\mu$ of the remnants, since such a calculation requires the cosmological context of the event and its progenitors, particularly the ages of the stellar populations. We therefore estimate $r_e$ and $\mu$ for progressively older stellar age values. Assuming that the progenitors formed at roughly the same early time, we initially set the age of the pre-existing (old) stars to 7~Gyr, while the age of stars formed during the merger is taken to be the elapsed simulation time (as if the simulation ended at $z=0$). We then repeat the calculation five times, increasing the age of all stars by 1~Gyr ($\Delta t$) at each step. These age increments implicitly assume that the remnants do not evolve dynamically until $z=0$, which is an approximation to secular, isolated evolution. For the metallicity, we adopt $Z=0.0004$ for the old stars, following local dwarfs of similar mass \citep{kirby2013universal}, and use the simulation values for the newly formed stars.

We compute $r_e$ and the mean surface brightness at this radius $\langle \mu\rangle_e$ by first assigning a light flux to each stellar particle based on their age and metallicity using the simple stellar population (SSP) models of \citet{10.1046/j.1365-8711.2003.06897.x}, in their 2016 version, which combines the MILES spectral libraries \citep{refId02} with the `Padova~1994' evolutionary tracks. The SSP models used assume a Kroupa IMF, consistent with our simulations. From them, we extract the flux in the CFHT $g$-band, which is commonly used in UDG studies. We then project the galaxies orienting the z-axis with the total angular momentum vector. We center the 1:1 remnant at the NSC, and the 1:2 at the stellar center of mass, and compute the cumulative flux profile. From it, we extract $r_e$ and calculate $\langle \mu\rangle_e$.

Table~\ref{tab:remnants_obs_prop} shows $r_e$ and $\langle \mu_g\rangle_e$ for the remnants at different $\Delta t$. For the youngest stellar ages ($\Delta t = 0$), the remnants are too compact and bright to qualify as UDGs. For $\Delta t \geq 2$~Gyr, corresponding to merger progenitor ages $\geq 9$~Gyr and mean ages of the newly formed stars $\geq 3.5$~Gyr (1:1 case) and $\geq 2.7$~Gyr (1:2 case), the remnants become extended enough to fall within the UDG regime ($r_e \gtrsim 1.5$~kpc). In terms of surface
brightness, UDGs have been defined using the effective surface brightness (that is, $\mu_e \geq 24$ and $\langle\mu\rangle_e \geq 24~\mathrm{mag\,arcsec^{-2}}$), rather than only the central surface brightness
\citep{di2017nihao, 2016A&A...590A..20V, van2022s}. As seen in Table~\ref{tab:remnants_obs_prop}, the remnant galaxies have sufficiently low surface brightness to lie within the UDG regime. The radii and surface
brightnesses in Table~\ref{tab:remnants_obs_prop} can be interpreted as an approximate time evolution of the remnants under the assumption of secular evolution: they transition from compact, bright galaxies into UDGs as their stellar populations age. As the newly formed stars age, the radial light-flux gradient flattens and
the light profile comes to approximate the underlying mass profile of the galaxies.

\begin{table}
\centering
\begin{tabular}{|c|c|c|c|c|}
\hline
 & \multicolumn{2}{c|}{1:1 remnant} & \multicolumn{2}{c|}{1:2 remnant} \\
\hline
        & $r_e$ & $\langle \mu_g\rangle_e$ & $r_e$ & $\langle \mu_g\rangle_e$  \\
{[Gyr]} & [kpc] & [mag\,arcsec$^{-2}$]  & [kpc] & [mag\,arcsec$^{-2}$]   \\
\hline
$\Delta t = 0$ & 0.16 & 20.5 & 0.39 & 22.2 \\
$\Delta t = 1$ & 1.19 & 25.6 & 1.45 & 25.8 \\
$2$            & 1.45 & 26.3 & 1.65 & 26.3 \\
$3$            & 1.55 & 26.6 & 1.71 & 26.5 \\
$4$            & 1.62 & 26.8 & 1.75 & 26.7 \\
$5$            & 1.66 & 26.9 & 1.78 & 26.8 \\
\hline
\end{tabular}
\caption{Half-light radii $r_e$ and mean surface brightness within this radius $\langle \mu_g \rangle_e$ of the remnants for different stellar ages. The bigger the $\Delta t$ value, the older the stellar population. Luminosities are computed in the g-band using \textit{GALAXEV} \citep{10.1046/j.1365-8711.2003.06897.x}. The remnants start as compact and bright sources and evolve into dim and large galaxies compatible with UDGs.}
\label{tab:remnants_obs_prop}
\end{table}

\subsection{The origin of UDGs}

To contextualize our idealized simulations and describe the implications of our work, we start by describing the progenitor galaxies. The progenitor galaxies are gas-rich UDGs themselves set by our initial conditions with stellar and gas masses that align with HI-selected UDGs and LSBs where the gas mass is greater than the stellar mass \citep{kado2022ultra, o2025wallaby}. Based on the two types of UDGs found by \citet{buzzo2025multiple}, our progenitors would belong to the "puffy dwarf" UDGs. These galaxies are characterized by having few or no GCs. Similarly, we find no GCC formation in the isolated UDG simulation despite having a large gas reservoir. The plausible origin for our progenitor puffy dwarf UDGs could be a regular dwarf galaxy that becomes more extended due to feedback expansion, formation within a high-spin halo, or tidal heating \citep{amorisco2016ultradiffuse, di2017nihao, carleton2019formation, yozin2015quenching}.

Our simulations present a formation pathway to turn puffy dwarf UDGs, which have no GCs, into GC-rich UDGs via mergers. The star clusters inhabiting the post-merger UDGs (referred to as GCCs) show similar radius, mass, density, and velocity dispersion as observed YMSCs and MW GCs (see Fig. \ref{fig:struct_gcs_props}), with the mean age of the GCCs being 1.5 Gyr in our longer simulation. Assuming that the GCCs evolve and become old GCs, which is plausible if the UDG has a secular evolution after the merger, the two post-merger UDGs would have $N_{GC}\gtrsim20$ and $M_{\rm GC}/M_{\star}\sim10\%$. Considering this and based on the two-type classification adopted in \citet{forbes2025some} and \citet{buzzo2025multiple} for UDGs, our post-merger galaxies would be part of the GC-rich population of UDGs. We note that the number of GCs can increase in a more realistic treatment where the progenitors host few GCs. Additionally, the post-merger UDGs show a cuspy DM density profile as the progenitors. Despite presenting the formation of GC-rich UDGs from progenitors that are UDGs themselves, we suggest that this formation pathway can be effective in mergers involving regular dwarfs (e.g. UDG-dwarf merger). In future work, we plan to study this formation pathway considering dwarfs in the progenitors.

Although $M_{\rm GC}/M_\star$ depends on the initial conditions we chose, we find that the mergers are an effective mechanism of significantly increasing the number of massive star clusters in the post-merger galaxy while only slightly increasing total stellar mass. This happens because most of the star formation takes place inside the clusters (see Fig. \ref{fig:scs_fraction}). This also preserves the faintness of the post-merger UDG, and solves the apparent contradiction between the intense star formation required to build massive star clusters and the mild star formation expected of a faint galaxy. Our previous numerical simulations of massive SC formation in interacting and merging galaxies showed that the formation efficiency ($M_{\rm GC}/M_\star$) of SCs can significantly increase due to the enhanced number of gas clouds with high gaseous pressure \citep{2002MNRAS.335.1176B}. However, the efficiency during interaction and merging between normal disk galaxies can be enhanced up to only 0.1, which is significantly lower than the values we find in this work for UDGs $M_{\rm SC}/M_{\star \rm new}>0.5$. This implies that the enhancement factor of massive SC formation efficiency might depend on the surface mass densities of the merger progenitor galaxies.

\subsection{GC-rich UDGs formed via mergers}

The specific properties of the UDGs formed via mergers depend on the merger parameters. In terms of GC-richness, the number of GCCs formed is sensitive to the collision orbit. For instance, in a 1:1 merger simulation in which we varied only the apocenter from 25 to 50 kpc, a few GCCs formed, but only one survived to the end of the simulation. The presence of galactic rotation depends on the progenitor masses and merger configuration: our 1:1 merger erases the initial rotation of the progenitors, whereas the 1:2 case retains some residual rotation. In contrast, the cuspy DM density profile of the progenitors is retained in the remnants of both simulations, indicating that this property is more robust to the initial conditions. Regarding the gas content, our post-merger UDGs retain only $\sim10\%$ of the initial gas mass, significantly lower than the trend followed by HI-selected UDGs and LSBs \citep{kado2022ultra, o2025wallaby, Leisman_2017}. In both runs, the radial distribution of GCCs is also more concentrated than the stellar distribution of the UDG, with the stars formed during the merger located preferentially in the inner regions of the galaxy. Finally, both runs follow the cluster number with halo mass relation.

UGC 9050-Dw1, described by \citet{fielder2023disturbed}, has been suggested to be a UDG formed by a dwarf merger, plausibly analogous to the model we present. This galaxy has a stellar mass similar to that of our post-merger UDGs. Although the gas mass is higher in the observed galaxy than in our model, this galaxy is in a low-density environment where the gas might have fallen back after the merger. We highlight the similarity between the core and surrounding diffuse components of UGC 9050-Dw1 and the two-component stellar density profile of our 1:1 remnant (see Fig.~\ref{fig:density_profile_stars11}): an inner component dominated by stars formed during the merger, and an outer component comprising the old stellar envelope of the progenitors. Analogous to the core of UGC 9050-Dw1, which is comparatively brighter and has a UV counterpart, the central part of our remnant is also brighter and actively forming stars. Concerning GCs, the observed galaxy shows that the majority of its GCs lie within its effective radius. Likewise, we report that the GCCs are centrally concentrated with $r_{h,\rm GC}<r_h$. Additionally, the observed GCs are monochromatic, a feature that our cluster candidates can reproduce, as they formed around the same age (during the galactic encounters) with a similar metallicity. We note, however, that our simulations do not produce as many star clusters as reported for UGC 9050-Dw1.

Our simulations can clarify the distinction between UDGs formed via interaction or full merger. \citet{2025A&A...699A..94B} report the detection of a rich population of young massive star clusters and old GCs in GAMA 526784, a gas-rich UDG in a low-density environment. Subsequently, \citet{2025A&A...700A.165B} conclude that this UDG can be the aftermath of a merger or an interaction, favoring the interaction hypothesis. The authors suggest that if a merger had taken place, the massive star cluster formation had to have happened many millions of years after the passages. Considering our simulations, we show that this cannot take place in a merger because most of the GCC formation occurs during the passages, and once the progenitors coalesce, the SFR is comparatively mild. Thus, we support the interaction formation pathway for GAMA 526784 rather than a merger.

\subsection{Globular cluster properties}

Although the simulations in this paper are specific to UDG formation, the GCC properties we find can be applied to more general systems containing massive SCs. The simulated star clusters show radii, velocity dispersion, density, and mass similar to those of MW GCs and YMSCs (see Fig.~\ref{fig:struct_gcs_props}). GCCs show metallicity spread values comparable to those observed in MW GCs (see Fig.~\ref{fig:metallicity_corr}, lower panel). Additionally, cluster metallicity is correlated with formation time (see Fig.~\ref{fig:metallicity_corr}, upper panel), similar to the age-metallicity relation observed in galaxies and star clusters, for instance in the LMC \citep{2022A&A...666A..80N, 1998MNRAS.296.1045C}. The specific angular momentum per radius $L/(Mr)$ is correlated with cluster mass, similar to the mass-$L/M$ correlation reported by \citet{2020ApJ...904...71L}. Similarly, we find that more massive clusters have progressively higher $V/\sigma$ values, as reported for MW GCs \citep{bianchini2018internal}. We note, however, that our simulated GCs differ from those formed in their own DM haloes (as shown in the simulations of \citealt{2024ApJ...971..103G, taylor2025emergence}), owing to their different formation mechanisms. These differences could be connected to the observed metal-rich and metal-poor GC populations \citep{2006ARA&A..44..193B}, with the merger-formed clusters linked to the metal-rich population and the halo-formed clusters to the metal-poor one.

\subsection{Limitations of this work}

Our simulations do not model the collisional two-body dynamics that drive the internal evolution of real star clusters. Each stellar particle represents a single stellar population rather than an individual star, and gravity is softened on a scale $\epsilon = 1\,\mathrm{pc}$; even in a softened system, however, the particle number remains the principal driver of the two-body relaxation timescale \citep{2008gady.book.....B}. The most direct consequence is that the sizes and densities of our GCCs are resolution-limited. In a convergence study of clusters formed in cosmological simulations, \citet{taylor2025emergence} show that, while cluster mass and metallicity are robust to spatial resolution, initial cluster sizes shrink as the resolution is increased; they further note that, in a full treatment, subsequent two-body expansion erases the memory of this initial radius. In a companion study using a collisional N-body code, \citet{taylor2025star} shows that resolution-limited simulations understate the survival rate of clusters because the limited resolution underestimates their birth density and renders them more susceptible to tidal disruption. The number of GCCs reported in this work can therefore be considered a lower limit, owing to our limited resolution.

Despite the mass and spatial resolution limitations just discussed, the structural parameters we compare with observations in Fig.~\ref{fig:struct_gcs_props} are set mainly at formation---by the compression of gas during the merger---rather than by subsequent collisional evolution. We confirm this by tracking the time evolution of the GCC properties: the mass and velocity dispersion are established at formation and remain relatively constant, whereas the radius grows,
plausibly through a combination of artificial (numerical) relaxation and physical tidal heating. The mass--velocity-dispersion relation reflects virial equilibrium and is preserved by the collisionless evolution, and this is the principal reason our clusters recover the observed scaling relations. Our simulations do not reproduce the internal structure of real GCs, which is shaped by mass segregation and core collapse driven by a realistic stellar mass spectrum that we do not model.

Pre-supernova feedback — radiation (photoionization and radiation pressure) and stellar winds — is an important regulator of the baryon cycle on both galactic and cluster scales. Its self-consistent inclusion lies beyond the scope of the present simulations, which adopt a thermal, IMF-averaged prescription for supernova feedback; the radiative and wind channels are not captured by this scheme, and the mass resolution adopted here (star-particle mass of $666.7~\rm M_\odot$ and gas-particle mass $2000\,\rm M_\odot$) only marginally resolves the dense interstellar structure that these channels most directly regulate. On galactic scales, the time-averaged star formation rate is relatively insensitive to the inclusion of early feedback \citep{2024A&A...681A..28A, 10.1093/mnras/stab1896}, so that the total stellar mass formed in our simulations — and hence the mass budget available to the remnants and their cluster populations — is not expected to change substantially. On star-cluster scales, by contrast, pre-SN feedback suppresses cluster formation at the high-mass end \citep{2024A&A...681A..28A, 2025A&A...704A.240D}; we therefore caution that its omission could artificially enhance the formation of massive star clusters in our simulations. However, mergers are widely identified as sites of massive star cluster formation, with such clusters regarded as the likely progenitors of GCs \citep{1992ApJ...384...50A, 10.1093/mnras/stw2969, 10.1093/mnras/staa1439}; we do not expect our GCCs to be purely artifacts of pre-SN feedback omission.

 \section{Conclusions}

We have presented a model for the formation of GC–rich ultra–diffuse galaxies (UDGs) triggered by gas–rich UDG–UDG mergers, using high–resolution, idealized SPH simulations. The high resolution adopted leads to resolving the formation of massive star clusters similar to GCs. Additionally, we have utilized a machine learning clustering algorithm, HDBSCAN, for the automatic detection of the GC candidates formed during the simulation. The main results concerning our two merger simulations and one isolated UDG are:

\begin{enumerate}
    \item The post-merger galaxies have properties similar to UDGs, large half-mass radius and low stellar mass. Observationally, the remnants evolve from compact, bright galaxies into UDGs in $\sim 2$ Gyr.
    \item Tidal compression and shocks during the galactic encounters of the merger drive efficient bound–cluster formation, yielding a high massive star cluster formation efficiency with \( M_{\mathrm{SC}}/M_{\star \mathrm{new}} \gtrsim 0.5 \).
  
    \item A significant portion of the massive star clusters (our GC candidates; GCCs) formed during the encounters survives to the end of the simulations, placing the remnant UDGs in the GC-rich regime. The 1:1 merger leads to 20 GCCs with the stellar mass of the post-merger galaxy locked in GCCs being 9$\%$, whereas the 1:2 simulation leads to 39 GCCs and 14$\%$ of the stellar mass locked in clusters. The mean age of the GCCs in the longer run is \(\sim 1.5~\mathrm{Gyr}\). The GCCs are centrally concentrated with a half-mass/half-number radius lower than the galaxy radius.
    \item The post-merger UDGs show weak or no rotation, they are dispersion-supported.
    \item The post-merger UDGs show a cuspy DM density profile as the progenitors.
    \item The GCCs occupy the same structural parameter space and follow similar empirical trends as YMSCs and MW GCs, including velocity–dispersion–mass, density–mass, and rotational velocity–mass relations, as well as chemical trends (metallicity–age and metallicity–dispersion–mass relations). These agreements support the interpretation of the simulated clusters as bona fide GC candidates.
    \item At the end of the merger simulations, the number of GCCs formed during the merger follows the GC number halo mass relation as has been observed in GC-rich UDGs.
    \item In one remnant, a nucleated UDG forms: the nuclear star cluster is initially seeded by infalling GCCs and subsequently grows via sustained continuous star formation.
    \item Although the isolated UDG is gas-rich, it does not form massive star clusters.
\end{enumerate}

In future work, we will extend this framework to:
\begin{itemize}
    \item Resolve the formation pathways of \emph{individual} GCCs (gas inflow, fragmentation, bound fraction, and feedback–regulated assembly);
    \item Follow the origin and growth of the nuclear star cluster (NSC), quantifying the roles of in–situ star formation versus GCC inspiral via dynamical friction;
    \item Compare the cluster formation efficiency \(M_{\mathrm{SC}}/M_{\star,\mathrm{new}}\) in dwarf–dwarf mergers against that in \emph{luminous} (more massive, high–SFR) galaxy mergers under similar conditions;
    \item Investigate the origin of cluster rotation, testing the imprint of natal angular momentum, tidal torques, and their scaling with mass and age.
\end{itemize}

Our simulations are idealized and do not include a cosmological environment; longer integrations and additional physics (e.g., alternative feedback channels, metal mixing, and an external tidal field) will further test the robustness of this channel and the long–term dynamical evolution of the GCC systems. Nonetheless, the results demonstrate that gas-rich mergers can provide a viable pathway to forming GC–rich---and in some cases nucleated---UDGs. 

\section*{Acknowledgements}

This work was supported by the University Postgraduate Award and the Scholarship for International Research Fees at The University of Western Australia.

This research was supported by the Australian Research Council (ARC) through Discovery Project DP220101863.

This work was supported by JSPS KAKENHI Grant Number 21K03621.

Numerical computations were carried out on Cray XC50 and Cray XD2000 at the Center for Computational Astrophysics (CfCA), National Astronomical Observatory of Japan. Numerical analysis was carried out in part on the analysis servers at CfCA.

Data analysis was partially performed on the OzSTAR national facility at Swinburne University of Technology. The OzSTAR program receives funding in part from the Astronomy National Collaborative Research Infrastructure Strategy (NCRIS) allocation provided by the Australian Government, and from the Victorian Higher Education State Investment Fund (VHESIF) provided by the Victorian Government.

We thank the referee for the insightful and constructive comments which improved this study.

%%%%%%%%%%%%%%%%%%%%%%%%%%%%%%%%%%%%%%%%%%%%%%%%%%
\section*{Data Availability}
 
The data used in this paper, including the merger and isolated simulations, will be shared on reasonable request to the corresponding author.



\bibliographystyle{mnras}
\bibliography{example} % if your bibtex file is called example.bib

%\begin{thebibliography}{99}
%\end{thebibliography}

%%%%%%%%%%%%%%%%%%%%%%%%%%%%%%%%%%%%%%%%%%%%%%%%%%

%%%%%%%%%%%%%%%%% APPENDICES %%%%%%%%%%%%%%%%%%%%%

\appendix

\section{Supplementary Galactic Properties}

\begin{figure}
\centering
\includegraphics[width=1\linewidth]{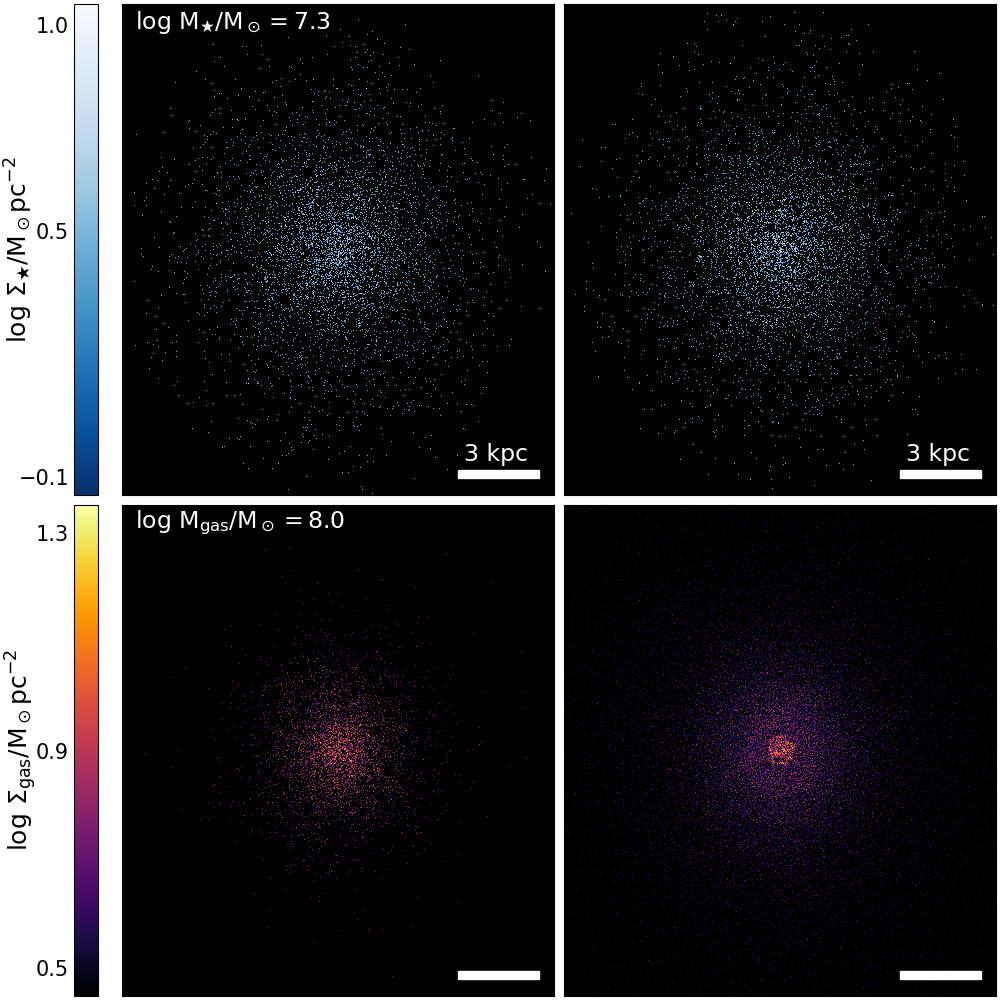}
\caption{Initial and final surface–density maps for the isolated progenitor \textbf{A} (or 1) run (galaxy properties in Table~\ref{tab:galaxies}). Top row: stellar; bottom row: gas. Columns show the initial conditions (left) and the final snapshot (right). The system evolves largely unperturbed and quiescent for 1.4 Gyr, with only modest gas redistribution; despite its high gas content, no GCCs form.}
\label{fig:snapshots_isolated}
\end{figure}

\begin{figure}
\centering
\includegraphics[width=1\linewidth]{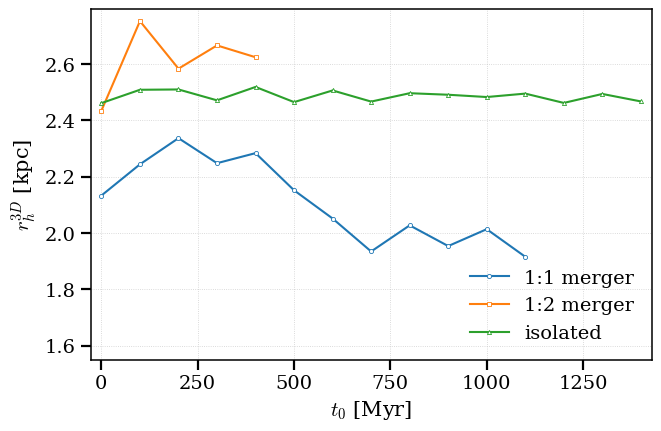}
\caption{Evolution of the three-dimensional stellar half-mass radius for the merger remnants and the isolated model (styles/colors as indicated in the legend). For the merger runs, time is measured relative to the coalescence epoch with $t_0=0$ being the coalescence time shown in Fig.~\ref{fig:SFH_mergers}. The isolated galaxy maintains a nearly constant radius during quiescent evolution. In the 1:1 run—integrated to later times—the radius contracts by 200 pc after coalescence and then remains extended at $\sim2$ kpc.}
\label{fig:radius_evolution}
\end{figure}

\begin{figure}
\centering
\includegraphics[width=1\linewidth]{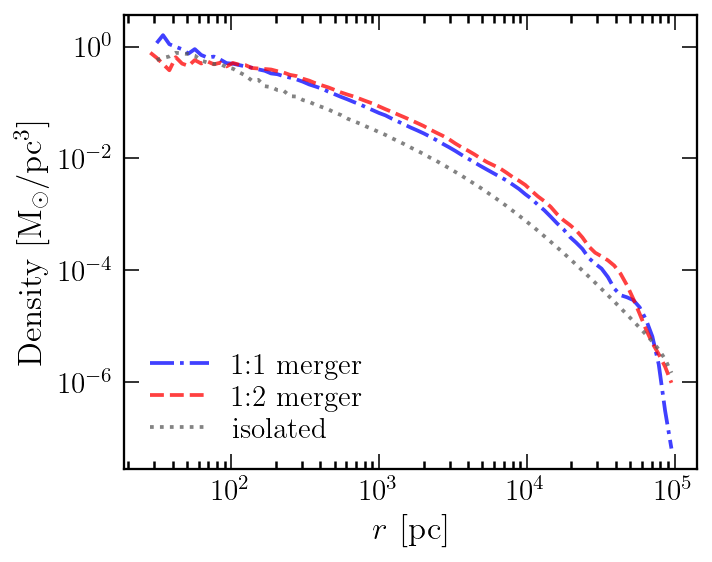}
\caption{
Spherically averaged dark-matter density profiles, as a function of 3D radius from the center of mass, for the 1:1 merger remnant (blue, dash-dotted), the 1:2 merger remnant (red, dashed), and the isolated galaxy
(grey, dotted). Both remnants are denser than the isolated model across the intermediate radii, and all three profiles truncate sharply near 100~kpc. Densities inside $\sim200$~pc lie below the convergence radius of \citet{2003MNRAS.338...14P} and are affected by two-body relaxation; they are excluded from our interpretation. 
}
\label{fig:dm_density_profiles}
\end{figure}

\begin{figure}
\centering
\includegraphics[width=1\linewidth]{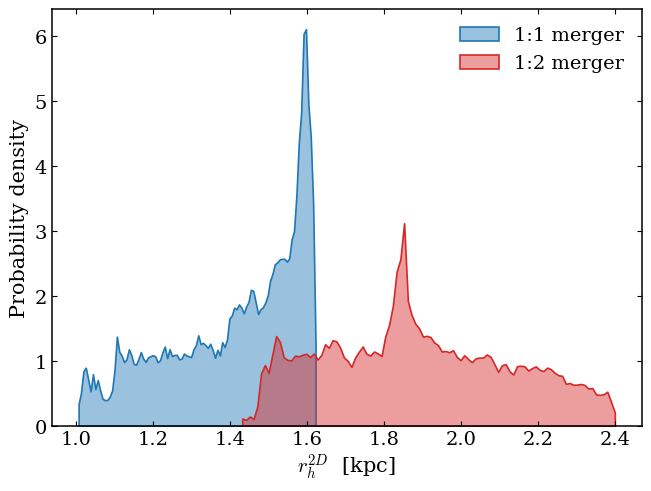}
\caption{Distributions of the projected half-mass radius for the merger remnants, computed over $2\times10^4$ random viewing angles and normalized to unity (styles/colors as indicated in the legend). The 1:2 remnant satisfies the UDG size criterion ($r\geq 1.5$ kpc) for nearly all viewing angles, whereas the 1:1 remnant exceeds this threshold for $\sim40\%$ of sightlines.}
\label{fig:r2D_distribution}
\end{figure}

In this section, we present supplementary figures concerning galactic properties of the simulations of this study. In Fig. \ref{fig:snapshots_isolated}, we show the initial and final snapshots of the isolated progenitor galaxy \textbf{A} in the control (isolated) run. This same model serves as the primary progenitor in both the 1:1 and 1:2 merger simulations. Over $\sim1.4$ Gyr, the system evolves quiescently, exhibiting only minor structural changes. Its 3D stellar half–mass radius remains approximately constant throughout the run as seen in Fig. \ref{fig:radius_evolution}, opposite to the variations seen in the remnant UDGs. In Fig.~\ref{fig:dm_density_profiles}, we show the DM density profiles of the three galaxies at the end of the simulations. Finally, Fig.~\ref{fig:r2D_distribution} shows the distribution of the projected (2D) half–mass radius, $r^{2D}_h$, for the post–merger remnants under random viewing angles.

\section{Supplementary Star Cluster Properties}
\label{app:scs_supplementary}

\begin{figure*}
\centering
\includegraphics[width=1\linewidth]{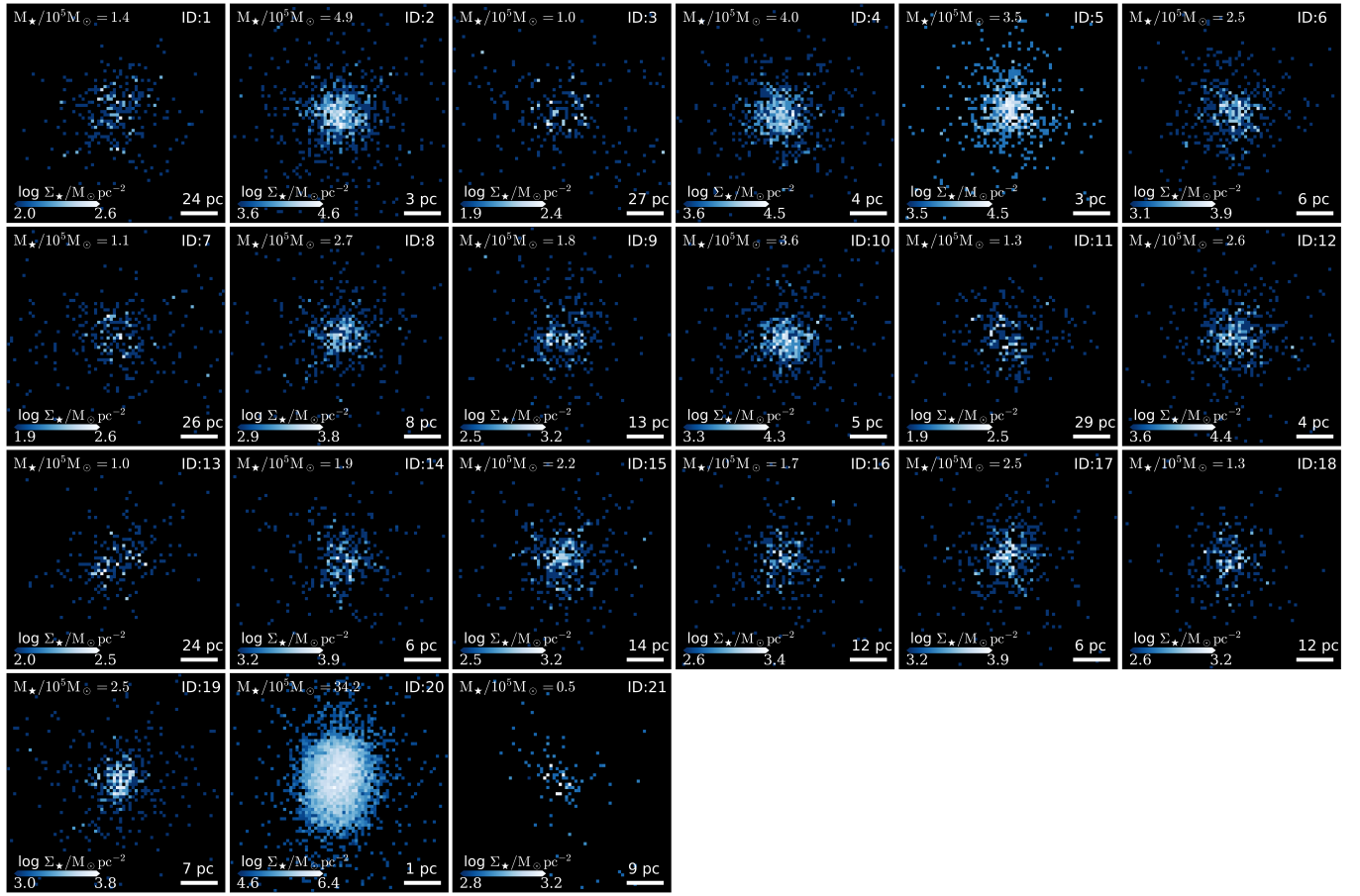}
\caption{Stellar surface–density maps of individual GCCs in the 1:1 post-merger remnant. Panels list each cluster’s stellar mass and ID; the IDs correspond to Table~\ref{tab:gcs11}. The cluster number 20 is an NSC.}
\label{fig:gcs11}
\end{figure*}

\begin{figure*}
\centering
\includegraphics[width=1\linewidth]{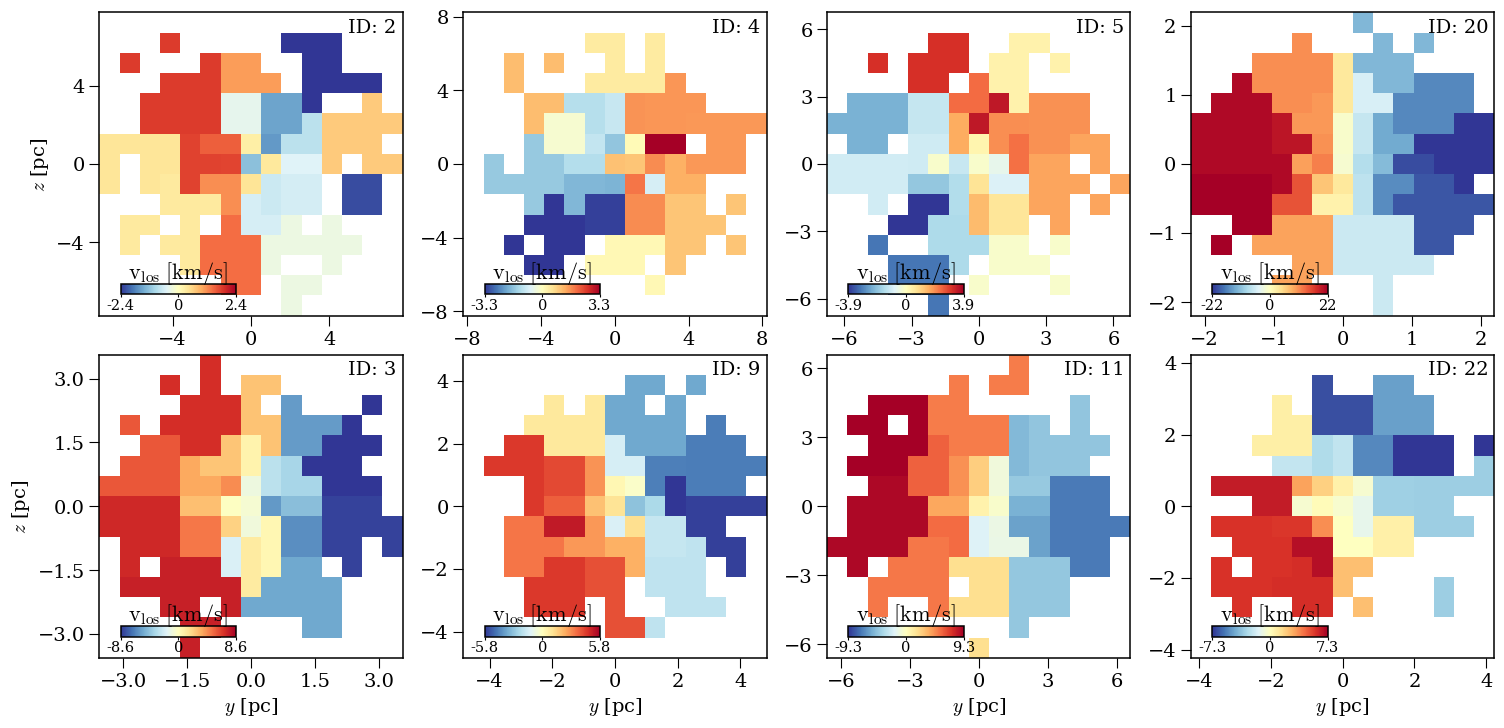}
\caption{Line-of-sight velocity maps for a sample of GCCs with high rotation in each post-merger remnant. Top row: clusters from the 1:1 galaxy; bottom row: GCCs from the 1:2 remnant. In all panels, the cluster total angular-momentum vector is aligned with the z-axis. Panel labels list cluster IDs matching Tables~\ref{tab:gcs11} and \ref{tab:gcs12}. Although the GCCs are dispersion-dominated ($V_\text{rot}/\sigma<1$), some nonetheless show clear signatures of ordered rotation in some cases. The maps adopt a 2D Voronoi binning using the PowerBin method by \citet{Cappellari2025}.}
\label{fig:gcs_los_velocity}
\end{figure*}

\begin{figure}
\centering
\includegraphics[width=1\linewidth]{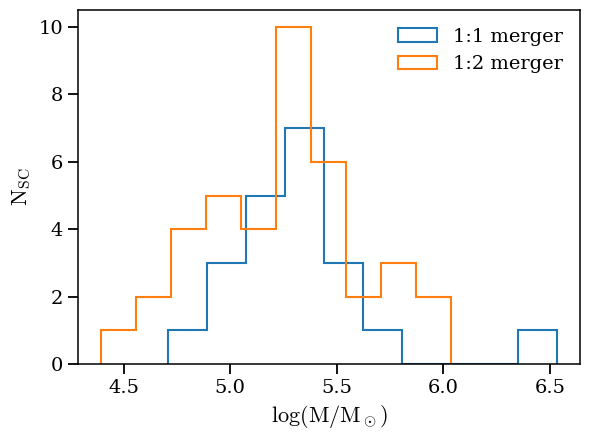}
\caption{Distribution of the number of star clusters across logarithmic mass bins. The star clusters are detected by our pipeline at the end of the merger simulations. The clusters include the GCCs plus the NSC formed in the 1:1 remnant. The two distributions show symmetry around a peak at nearly the same mass, which agrees with the universality of the GCLF.}
\label{fig:gcmf}
\end{figure}

In this section, we present supplementary figures concerning GCC properties. In Figures \ref{fig:gcs11} and \ref{fig:gcs12}, we present stellar surface–density maps of the GCCs identified in the merger remnants; their properties are shown in Tables \ref{tab:gcs11} and \ref{tab:gcs12}. We show line–of–sight velocity maps for the GCCs with the strongest rotation in each merger remnant in Fig.~\ref{fig:gcs_los_velocity}. Finally, we show the mass distribution in Fig.~\ref{fig:gcmf}.

\begin{figure*}
\centering
\includegraphics[width=1\linewidth]{gcs12.png}
\caption{Stellar surface–density maps of individual GCCs in the 1:2 post-merger remnant. Panels list each cluster’s stellar mass and ID; the IDs correspond to Table~\ref{tab:gcs12}.}
\label{fig:gcs12}
\end{figure*}

\begin{table*}
\caption{Properties of the GCCs in the 1:1 post-merger remnant. Corresponding surface-density maps are shown in Fig.~\ref{fig:gcs11}. Columns (left to right): cluster ID, mass, 3D half-mass radius $r^{3D}_{h}$, density within half-mass radius $\rho_{h}$, 1D velocity dispersion $\sigma$, mean tangential rotational velocity $V_\text{rot}$, specific angular momentum $\rm L/\rm M$, mean age, age standard deviation $\sigma_{\rm age}$, mean iron-to-hydrogen ratio [Fe/H], iron-to-hydrogen standard deviation $\sigma_{[\mathrm{Fe}/\mathrm{H}]}$, and distance to the galactic center $D_{\rm gal}$. We include the NSC as the cluster number 20.}
\label{tab:gcs11}
\begin{tabular}{lccccccccccc}
\toprule

ID & Mass             & $r^{3D}_{h}$ & $\rho_{h}$           & $\sigma$ & $V_\text{rot}$ & L/M                     & $age$           & $\sigma_{\rm age}$ & $[\mathrm{Fe}/\mathrm{H}]$ & $\sigma_{[\mathrm{Fe}/\mathrm{H}]}$ & $D_{\rm gal}$ \\

&[$10^5\rm M_\odot$]  & [pc]         & [$\rm M_\odot/pc^3$] & [km/s]   & [km/s]         & [$\rm km~\rm pc/\rm s$] & [Gyr]           & [Myr]          &                            & $(\times10^{-2})$                   & [kpc] \\
   
\midrule
1 & 1.4 & 24 & 1.2 & 1.9 & 0.14 & 2.5 & 1.3 & 1.2 & -1.2 & 2.0 & 3.7 \\
2 & 4.9 & 3.9 & 999 & 7.4 & 1.5 & 6.5 & 1.7 & 2.2 & -1.4 & 6.6 & 3.3 \\
3 & 0.97 & 26 & 0.65 & 2.9 & 0.41 & 16 & 1.3 & 2.2 & -1.2 & 4.1 & 2.5 \\
4 & 4.0 & 4.1 & 689 & 6.5 & 1.7 & 7.4 & 1.7 & 1.9 & -1.4 & 3.2 & 2.8 \\
5 & 3.5 & 3.4 & 1090 & 6.5 & 1.8 & 6.8 & 1.3 & 1.8 & -1.1 & 5.0 & 2.3 \\
6 & 2.5 & 6.8 & 95 & 4.4 & 0.47 & 3.5 & 1.7 & 2.0 & -1.4 & 5.6 & 2.1 \\
7 & 1.1 & 27 & 0.66 & 1.9 & 0.070 & 3.4 & 1.7 & 1.7 & -1.4 & 2.2 & 3.4 \\
8 & 2.7 & 8.7 & 49 & 4.1 & 0.38 & 4.0 & 1.7 & 1.3 & -1.4 & 0.51 & 1.2 \\
9 & 1.8 & 14 & 7.8 & 2.8 & 0.29 & 8.1 & 1.5 & 1.8 & -1.3 & 2.2 & 0.80 \\
10 & 3.6 & 5.3 & 285 & 5.7 & 0.68 & 4.0 & 1.7 & 1.9 & -1.4 & 6.4 & 2.0 \\
11 & 1.3 & 30 & 0.59 & 1.8 & 0.26 & 9.1 & 1.6 & 1.9 & -1.4 & 1.2 & 0.95 \\
12 & 2.6 & 4.1 & 464 & 5.3 & 0.79 & 3.7 & 1.3 & 2.0 & -1.1 & 1.6 & 1.3 \\
13 & 1.0 & 24 & 0.85 & 1.7 & 0.19 & 6.3 & 1.5 & 1.5 & -1.4 & 0.74 & 1.1 \\
14 & 1.9 & 6.7 & 77 & 3.8 & 0.37 & 2.2 & 1.3 & 1.7 & -1.2 & 2.7 & 1.6 \\
15 & 2.2 & 14 & 8.5 & 3.2 & 0.20 & 2.3 & 1.7 & 2.0 & -1.4 & 3.3 & 0.77 \\
16 & 1.7 & 12 & 11 & 2.9 & 0.34 & 4.8 & 1.5 & 1.1 & -1.3 & 0.64 & 0.41 \\
17 & 2.5 & 6.7 & 100 & 4.3 & 0.73 & 4.7 & 1.7 & 1.7 & -1.4 & 1.5 & 0.48 \\
18 & 1.3 & 13 & 7.5 & 2.6 & 0.059 & 1.5 & 1.5 & 2.2 & -1.3 & 6.1 & 0.65 \\
19 & 2.5 & 7.9 & 60 & 4.3 & 0.71 & 4.5 & 1.6 & 1.8 & -1.4 & 1.2 & 0.50 \\
20 & 34 & 1.1 & $3.1 \times 10^{5}$ & 26 & 18 & 19 & 0.61 & 482 & -0.38 & 57 & 0.033 \\
21 & 0.51 & 9.4 & 7.3 & 2.5 & 0.35 & 5.3 & 0.69 & 70 & -0.74 & 5.3 & 0.14 \\
\hline
\textbf{Median} & 2.2 & 8.7 & 48.7 & 3.8 & 0.4 & 4.7 & 1.5 & 1.9 & -1.4 & 2.7 & 1.2\\
\bottomrule
\end{tabular}
\end{table*}

\begin{table*}
\caption{Properties of the GCCs in the 1:2 post-merger remnant. Columns follow the caption in Table \ref{tab:gcs11}. Corresponding surface-density maps are shown in Fig.~\ref{fig:gcs12}.}
\label{tab:gcs12}
\begin{tabular}{lccccccccccc}
\toprule

ID & Mass             & $r^{3D}_{h}$ & $\rho_{h}$           & $\sigma$ & $V_\text{rot}$ & L/M                     & $age$           & $\sigma_{\rm age}$ & $[\mathrm{Fe}/\mathrm{H}]$ & $\sigma_{[\mathrm{Fe}/\mathrm{H}]}$ & $D_{\rm gal}$ \\

&[$10^5\rm M_\odot$]  & [pc]         & [$\rm M_\odot/pc^3$] & [km/s]   & [km/s]         & [$\rm km~\rm pc/\rm s$] & [Gyr]           & [Myr]          &                            & $(\times10^{-2})$                   & [kpc] \\
   
\midrule
1 & 1.9 & 3.7 & 444 & 4.8 & 0.55 & 2.2 & 0.74 & 1.8 & -1.4 & 1.1 & 5.2 \\
2 & 6.0 & 2.1 & 7427 & 9.3 & 4.2 & 9.6 & 0.85 & 2.5 & -1.4 & 13 & 3.2 \\
3 & 9.5 & 1.8 & $2.0 \times 10^{4}$ & 12 & 5.6 & 12 & 0.84 & 3.0 & -1.3 & 20 & 4.4 \\
4 & 2.0 & 3.1 & 804 & 5.3 & 1.3 & 4.1 & 0.74 & 1.8 & -1.4 & 1.4 & 4.4 \\
5 & 3.4 & 2.9 & 1659 & 6.6 & 2.5 & 7.2 & 0.79 & 2.0 & -1.4 & 4.3 & 3.4 \\
6 & 0.90 & 23 & 0.90 & 1.6 & 0.36 & 5.7 & 0.75 & 2.1 & -1.4 & 2.2 & 4.5 \\
7 & 0.48 & 31 & 0.20 & 2.6 & 0.43 & 26 & 0.84 & 1.1 & -1.4 & 0.44 & 4.6 \\
8 & 0.95 & 23 & 0.89 & 1.9 & 0.24 & 10 & 0.65 & 2.0 & -1.3 & 2.9 & 1.9 \\
9 & 6.3 & 2.4 & 5325 & 9.3 & 5.1 & 13 & 0.85 & 2.3 & -1.4 & 9.8 & 0.38 \\
10 & 1.2 & 22 & 1.3 & 1.9 & 0.36 & 9.9 & 0.86 & 1.8 & -1.4 & 3.8 & 1.5 \\
11 & 11 & 3.3 & 3640 & 11 & 6.0 & 22 & 0.87 & 2.8 & -1.3 & 14 & 2.7 \\
12 & 0.25 & 36 & 0.065 & 15 & 4.4 & 144 & 0.58 & 45 & -1.3 & 3.6 & 1.1 \\
13 & 0.62 & 25 & 0.48 & 1.3 & 0.11 & 3.9 & 0.56 & 1.6 & -1.3 & 1.0 & 0.78 \\
14 & 3.8 & 2.2 & 4513 & 7.4 & 1.1 & 2.6 & 0.69 & 2.5 & -1.4 & 6.0 & 2.0 \\
15 & 2.1 & 4.0 & 384 & 4.7 & 1.5 & 7.6 & 0.47 & 2.0 & -1.3 & 2.7 & 1.7 \\
16 & 2.4 & 1.8 & 4513 & 6.2 & 3.2 & 7.2 & 0.48 & 1.8 & -1.3 & 1.2 & 0.48 \\
17 & 0.69 & 23 & 0.65 & 1.5 & 0.12 & 2.6 & 0.58 & 2.0 & -1.3 & 1.4 & 1.2 \\
18 & 3.0 & 1.7 & 7196 & 6.9 & 1.3 & 2.5 & 0.56 & 1.9 & -1.3 & 1.4 & 0.70 \\
19 & 3.4 & 1.9 & 5874 & 7.0 & 3.0 & 6.7 & 0.52 & 2.5 & -1.2 & 7.7 & 1.1 \\
20 & 1.7 & 8.1 & 38 & 3.4 & 1.00 & 8.6 & 0.68 & 2.0 & -1.4 & 4.1 & 0.46 \\
21 & 1.1 & 4.5 & 138 & 3.4 & -0.011 & 0.82 & 0.19 & 2.8 & -1.00 & 2.2 & 1.0 \\
22 & 4.8 & 2.1 & 6106 & 8.4 & 4.4 & 9.5 & 0.52 & 2.6 & -1.2 & 6.9 & 1.1 \\
23 & 0.72 & 17 & 1.7 & 1.8 & 0.061 & 2.6 & 0.51 & 1.5 & -1.2 & 1.2 & 0.65 \\
24 & 1.8 & 3.5 & 474 & 4.7 & 1.2 & 4.6 & 0.66 & 1.8 & -1.3 & 3.0 & 0.58 \\
25 & 0.52 & 14 & 2.1 & 2.0 & 0.13 & 4.0 & 0.23 & 2.6 & -0.99 & 2.3 & 0.31 \\
26 & 2.5 & 2.0 & 3675 & 6.0 & 0.93 & 3.4 & 0.65 & 1.6 & -1.3 & 4.2 & 0.67 \\
27 & 2.0 & 3.5 & 584 & 5.1 & 0.96 & 3.7 & 0.65 & 2.2 & -1.3 & 3.6 & 0.89 \\
28 & 1.1 & 15 & 4.3 & 2.3 & 0.055 & 2.1 & 0.67 & 1.4 & -1.3 & 3.4 & 0.88 \\
29 & 0.66 & 19 & 1.1 & 1.6 & 0.15 & 7.6 & 0.66 & 2.1 & -1.3 & 3.8 & 0.60 \\
30 & 0.97 & 6.9 & 35 & 2.9 & 1.1 & 8.7 & 0.18 & 2.8 & -1.0 & 3.6 & 0.37 \\
31 & 1.4 & 3.4 & 427 & 4.1 & 0.89 & 2.8 & 0.64 & 1.1 & -1.3 & 1.0 & 0.97 \\
32 & 2.5 & 2.2 & 2637 & 5.8 & 0.88 & 2.3 & 0.78 & 1.5 & -1.4 & 1.8 & 0.24 \\
33 & 2.1 & 0.97 & $2.7 \times 10^{4}$ & 6.3 & 1.1 & 1.0 & 0.24 & 2.0 & -0.99 & 2.8 & 0.76 \\
34 & 5.3 & 2.7 & 3083 & 8.6 & 3.1 & 9.0 & 0.67 & 43 & -1.4 & 5.8 & 0.90 \\
35 & 0.84 & 11 & 7.0 & 2.2 & 0.52 & 5.2 & 0.16 & 2.5 & -0.95 & 1.9 & 0.94 \\
36 & 1.2 & 0.85 & $2.4 \times 10^{4}$ & 5.1 & 3.0 & 2.5 & 0.13 & 2.0 & -0.92 & 2.2 & 0.43 \\
37 & 1.9 & 3.5 & 538 & 4.8 & 1.2 & 4.0 & 0.67 & 1.9 & -1.3 & 3.1 & 0.17 \\
38 & 1.9 & 0.89 & $3.3 \times 10^{4}$ & 5.7 & 3.8 & 3.5 & 0.13 & 1.8 & -0.92 & 2.0 & 0.55 \\
39 & 3.0 & 4.5 & 383 & 6.7 & 1.3 & 7.1 & 0.076 & 42 & -0.90 & 3.9 & 0.11 \\
\hline
\textbf{Median} & 1.9 & 3.5 & 473.6 & 5.1 & 1.1 & 5.2 & 0.7 & 2 & -1.3 & 3.03 & 0.9\\
\bottomrule
\end{tabular}
\end{table*}

%%%%%%%%%%%%%%%%%%%%%%%%%%%%%%%%%%%%%%%%%%%%%%%%%%

% Don't change these lines
\bsp	% typesetting comment
\label{lastpage}
\end{document}